\documentclass[12pt,reprint,onecolumn,aps,groupedaddress,nofootinbib,floatfix,showkeys]{revtex4-2}
\usepackage{amsmath,amsfonts,amssymb}
\usepackage{framed}
\usepackage{fourier}
\usepackage{graphicx,color}
\usepackage{hyperref}
\usepackage{bm}
\usepackage{multirow}
\usepackage{hyperref}
\usepackage{mathrsfs}
\usepackage{slashed}
\usepackage[capitalise]{cleveref}
\usepackage{cancel}
\usepackage{soul}
\usepackage{natbib}
\usepackage{mciteplus}
\usepackage{placeins}
\usepackage{lineno}
\usepackage{euscript}
\DeclareMathOperator{\sgn}{sgn}
 \usepackage[normalem]{ulem}

 \definecolor{darkgreen}{cmyk}{1,0,1,0.4}
 
\definecolor{ultramarine}{rgb}{0.5, 0.125, 0.376}

\def\bra{\langle}
\def\ket{\rangle}

\begin{document}

\title{Indications for new scalar resonances at the LHC and a possible interpretation}
\author{Anirban Kundu}
\email{akphy@caluniv.ac.in}
\affiliation{Department of Physics, University of Calcutta,
92 Acharya Prafulla Chandra Road, Kolkata 700009, India}
\author{Poulami Mondal}
\email{poulami.mondal@tifr.res.in} 
\affiliation{Department of Physics, Indian Institute of Technology Kanpur, 
Kanpur 208016, India, and\\
Department of Theoretical Physics, Tata Institute of Fundamental Research, 
Homi Bhabha Road, Mumbai 400005, India}
\author{Gilbert Moultaka}
\email{gilbert.moultaka@umontpellier.fr}
\affiliation{Laboratoire Univers \& Particules de Montpellier (LUPM), 
Universit\'e de Montpellier, CNRS, Montpellier, France}

\begin{abstract}
Over the last few years, the CMS and ATLAS collaborations at the Large Hadron Collider (LHC) have reported excesses that could hint at several new scalar resonances. Although none of them has touched the discovery level, at least two of them, at about 95 GeV and 650 GeV, have been indicated by more than one experiments, and have reached statistical significance worthy of a serious investigation. Conservatively using only the numbers given by the experimental collaborations, we find combined global significances around 3$\sigma$ and 4$\sigma$ respectively for the 95~GeV and 650~GeV putative resonances. There are some more, like the one at 320 GeV, which have also been hinted at. We show that the data on only the 650 GeV resonance, assuming they stand the test of time, predict the existence of a doubly-charged scalar, and make the more common extensions of the scalar sector like those by gauge singlet scalars, the 2-Higgs doublet models or the Georgi-Machacek model, highly  disfavored. We provide the readers with a minimalistic model that may possibly explain all the indications. Such a model can also accommodate the hints of a singly charged scalar at about 375 GeV, and a doubly charged scalar at about 450 GeV, as found by both the major LHC Collaborations, the combined global significance for each of them being above $2.5\sigma$. We show that even the scant data, with large error bars, have the potential to strongly constrain our model containing four scalar multiplets, which makes the model easily testable and falsifiable. Our analysis comes with the obvious caveat that the allowed parameter space that we find depends on the available data on all the new resonances, and may change in future. One may also note that this is an exploratory exercise that illustrates the difficulties when it comes to fitting several resonances simultaneously, even for next-to-minimal extensions of the SM.
\end{abstract}

\date{\today}
\maketitle

\tableofcontents

\section{Introduction} \label{sec:intro}
Twelve years after the discovery of the Higgs boson \cite{ATLAS:2012yve, CMS:2012qbp}, 
some more interesting hints about new resonances have been found
by both the ATLAS and the CMS Collaborations 
\cite{ATLAS:2017tlw,CMS:2017jkd, CMS:2018mmw, CMS:2015ocq, CMS:2018cyk, CMS:2022rbd, CMS:2022goy,ATLAS:2020tlo, ATLAS:2021kog, CMS:2022bcb, CMS:2022tgk, CMS:2019pzc, ATLAS:2022wti, ATLAS:2023zkt} 
at the Large Hadron Collider (LHC). 
It is way too early to say whether these hints will stand the test of time, but they surely deserve a close scrutiny. 
There have already been several such attempts in the context of specific models, {\em e.g.}, in Refs.\ 
\cite{Cea:2018tmm, Biekotter:2022jyr, poulami1, Crivellin:2021ubm,Iguro:2022dok,Banik:2023ecr}, 
although hardly any analysis considers all the indications simultaneously.

In this paper, we will deal with all the reasonable hints or indications of such scalar resonances. By reasonable, we mean 
only those resonances for which the combined global significance, as opposed to local significance 
which does not take into account the look-elsewhere effect (LEE), is at the level of $2.5\sigma$ or higher. 
In most of the cases these channels come from independent final states, and therefore the combined 
significance is simply obtained through a product of probabilities. 
Special emphasis will be given to those resonances that 
have been indicated by both the major LHC 
collaborations, or maybe by even earlier experiments. A preliminary overview of the scenario, including the signals and 
their significances,  may be found in Ref.\ \cite{Kundu:2022bpy}; however, in the present paper, we have been more 
conservative about the significance estimates. A comprehensive list of the hints is given in Section 
\ref{sec:signals}, and Table \ref{tab:significance}, of this paper. 
We will assume all resonances to be of spin-0 nature. They have been found to have nonzero coupling with the
Standard Model (SM) gauge bosons, so the mass eigenstates can hardly be dominated by gauge singlets. 
We will show that they cannot come from 
simple multi-Higgs doublet extensions either. In fact, the apparent indications tend to rule out, or at least 
disfavour, a number of minimal extensions of the scalar sector of the SM.

There are three major constraints on any such scalar extension. The first one is the existence of custodial
symmetry (CS), parametrised by the $\rho$-parameter. In the SM, $\rho=1$ at the tree-level, but 
the Yukawa interaction in the SM breaks the CS. Once the radiative corrections are incorporated, the $\rho$-parameter
shifts from unity, the dominant contribution coming from the top quark
loop and being proportional to $m_t^2$. To absorb this correction, one defines the parameter
$\rho_0$, so that in the SM, $\rho_0$ is unity by definition. The experimental value is close to
unity~\cite{ParticleDataGroup:2024cfk}:
\begin{eqnarray}
\rho_0 = 1.00031 \pm 0.00019\,,
\label{eq:rho-expt}
\end{eqnarray}
so CS must be close to an exact symmetry of nature. Models with SU(2) singlet and doublet scalar multiplets 
respect CS; it is also possible to include larger multiplets of scalars in a way that CS is
respected.  The most famous example of this genre is the Georgi-Machacek (GM) model \cite{Georgi:1985nv}, 
where a complex triplet and a real triplet of scalars engender $\rho=1$ at the tree-level\footnote{The CS is 
respected in the GM model even at the one-loop level if one considers only the SU(2) part of the weak interaction. 
Once $\rm U(1)_Y$ is introduced and gauged, the tadpole diagrams break the CS.}. Scalar extensions that 
violate CS must have very small vacuum expectation values (VEV) of the CS-breaking fields; thus, they are unlikely
to be singly produced by gauge boson fusion, or decay to the gauge bosons with any significant branching ratio (BR), 
as the (neutral) scalar-gauge-gauge coupling is always proportional to the VEV of that scalar. 

The GM model is interesting in the sense that this is the minimal model that preserves the CS and introduces both
singly and doubly charged scalars in the spectrum. One may even construct a more generalised version of the
GM model \cite{KMP22} where the CS is respected in the gauge sector only but not in the scalar sector, {\em i.e.},
the components of the custodial scalar multiplets can potentially be non-degenerate. This opens up more 
decay channels for the scalars. A comprehensive analysis of this model has been done recently
\cite{Mondal:2022xdy,Chowdhury:2024mfu,Mondal:2025tzi}. We stress this point because this will be important in our subsequent discussions. 

The second major constraint is the nature of the scalar resonance at 125 GeV, which we call $h_{125}$.
From the measurement of its couplings to the fermions and the
gauge bosons, we know that its properties must be very close to that of the SM Higgs boson $h_{\rm SM}$
\cite{CMSnature,toro}. Denoting the generic multiplicative modifier to the coupling of any particle-antiparticle pair
with the SM Higgs boson by $\kappa$, we have \cite{CMSnature}
\begin{equation}
\kappa_W = 1.02\pm 0.08\,,\ \ \kappa_Z = 1.04\pm 0.07\,,\ \ 
\kappa_t = 1.01^{+0.11}_{-0.10}\,,\ \ \kappa_b = 0.99^{+0.17}_{-0.16}\,. \label{eq:CMS-SMkappas}
\end{equation}
At this point, one may note that while $h_{125}\to\gamma\gamma$ enforces $\kappa_t$ and $\kappa_W$
to have the same sign, the sign of $\kappa_b$ is yet to be determined.
ATLAS and CMS Collaborations \cite{ATLASWh, CMSWh} showed, from the analysis of $Wh_{125}$ production
in the vector boson fusion (VBF) channel, that $\kappa_W = -\kappa_Z = 1$ is ruled out
with significance greater than $8\sigma$. Such conclusions, however,
strongly rely on the underlying assumption that
the processes under investigation are mediated by the SM particles only. It was shown in
Ref.\ \cite{dasPLB} that such a possibility can still be entertained if there is new physics
below at about 600 GeV. However, to be conservative, we will consider all $\kappa$s for $h_{125}$
to be positive and consistent with \cref{eq:CMS-SMkappas}.

The third major constraint is the preservation of unitarity in gauge boson scattering processes, like $W_LW_L
\to W_LW_L, Z_LZ_L$, where $L$ denotes the longitudinally polarised state. There are important sum rules that
must be valid for any extensions with elementary scalars for the underlying gauge theory to remain 
renormalisable and for the cross-section to fall
at high energies \cite{Gunion:1990kf}. For example, if the $h_{125}W^+W^-$ coupling is found to be exactly what is
expected in the SM, no other neutral scalar can couple to $W^+W^-$ unless there is at least one doubly charged
scalar in the theory. As this will play an important role in analysing the signal and developing our
model, we discuss this in a separate subsection.

Two of the strongest hints, as discussed in Ref.\ \cite{Kundu:2022bpy}, 
are the resonances roughly peaked at 95 GeV
 and 650 GeV, which we will call $h_{95}$ and $H_{650}$ respectively\footnote{We will generically use $h$ for 
 scalars whose masses are equal to or smaller than 125 GeV, and $H$ for the heavier ones.}. 
 While $h_{95}$ has received ample attention in the literature, $H_{650}$ has been relatively overlooked, a strange fact 
 considering that $H_{650}$ has, as we will see, the strongest combined global significance, above $4\sigma$.
 The former, $h_{95}$, is a narrow resonance while the 
 latter, $H_{650}$, is a broad one. CMS observed $h_{95}$ decaying to $\gamma\gamma$ and $\tau^+\tau^-$ while
 LEP found a hint of $h_{95}$, subsequently decaying to $b\bar{b}$, through $e^+e^-\to Z^*\to Zh_{95}$. 
 What we mean by observation is actually an excess in 
 the respective channels, which we have discussed later. There have been attempts to interpret this 
 resonance as a standalone gauge singlet scalar \cite{poulami1}, or a gauge singlet along with a second Higgs
 doublet \cite{Biekotter:2022jyr}. 
   
 Both ATLAS and CMS found an excess at about 650 GeV \cite{ATLAS:2020tlo,ATLAS:2021kog,CMS:2022bcb,
 CMS:2022tgk,ATLAS:2023zkt}, the 
 new resonance decaying mostly to $W^+W^-$ and also to $ZZ$. It also decays to $h_{95}h_{125}$. What is 
 important is the fact that in almost every channel, the local significance for $H_{650}$ 
 is more than $3\sigma$, as can be seen from Table~\ref{tab:significance}. 
 According to CMS, VBF is the preferred channel for  $H_{650}$ production,
 but whether it can also be produced through the gluon-gluon fusion (ggF) channel
 is still uncertain. Thus, whether $H_{650}$
 has a fermionic coupling is an open issue. We will see that its interpretation poses severe problems within 
 the popular extensions of the scalar sector of the SM, and will discuss a minimalistic model where this resonance 
 may be accommodated. 
  
There are more hints, most prominent among them are one at 400 GeV and another at about 320 GeV. While we are 
still very far from determining the CP properties of these bumps, the pattern indicates that $h_{95}$, $H_{650}$, 
and $H_{320}$, along with $h_{125}$, are CP-even, while $A_{400}$ is CP-odd, assuming that all of them are
CP-eigenstates. Obviously, they need not be, and that opens up more interesting possibilities. 
For completeness, 
we should also mention the scalar $A_{151}$, tentatively CP-odd, as surmised by the authors of Ref.\ 
\cite{Crivellin:2021ubm}
and further explored in Refs.\ \cite{Coloretti:2023wng, Bhattacharya:2023lmu},
through an analysis of the CMS and ATLAS data. There are also two mild hints of 
charged scalars, a singly charged scalar $H^+$ at about 375 GeV and a doubly charged scalar $H^{++}$ at about 450 
GeV \cite{CMS:2021wlt, ATLAS:2022zuc, ATLAS:2023dbw}. They will be discussed in more detail later.

We will show that the coupling of $H_{650}$ to $W^+W^-$ violates the unitarity sum rules
\cite{Gunion:1990kf} if $H_{650}$ is a part of 
an extended scalar sector consisting of only SU(2) singlets and doublets. 
A similar conclusion has been reached by the authors of Ref.\ \cite{LeYaouanc:2023zvi}.
In other words, existence of at least one
doubly charged scalar is needed. Complex triplets include doubly charged scalar fields, but in general they do not
respect the custodial symmetry; the Georgi-Machacek (GM) model, and its extensions,
being possible exceptions. We will, however, show that even the canonical GM model
\cite{Georgi:1985nv} falls short
for this; it can only partially explain the scalar spectrum. Thus, one needs a further extension to 
accommodate all the resonances. We will propose such a model here, without going into any details of the 
corresponding scalar potential, by augmenting the generalized GM model proposed 
in Ref.\ \cite{KMP22} with another SU(2) doublet. 
We call this model the 2-Higgs Doublet extended Georgi-Machacek (2HDeGM) model.
We emphasise that this is a minimal model, but highly predictive and restrictive, and hence easily falsifiable. 

The paper is organised as follows. In Section \ref{sec:signals}, we enlist all the interesting resonances that we talk about,
and show why they cannot possibly be explained in typical extensions like the multi-Higgs doublet models, or the 
canonical GM model. In Section \ref{sec:theory}, we propose 2HDeGM,
an extension of the GM model 
with an extra scalar SU(2) doublet, and show that this may satisfactorily explain all the existing hints without being 
in conflict with any existing data. In particular, we show that, even with whatever little data we have, and without 
any knowledge of the full scalar potential, we can infer a lot and severely constrain the parameter space of the model, and also make 
useful and testable predictions. We would like to stress again that 2HDeGM is a minimal type of model to accommodate
all the hints and their production and decay patterns as far as possible, but we will not venture into a study 
of the exact scalar potential and hence the spectrum,  necessitating involved stability and 
unitarity constraints, similar to \cite{Moultaka:2020dmb, Chowdhury:2024mfu}, as this is not even 
required to establish the central theme of the paper. 
In short, we never claim that even if all the signals persist, 2HDeGM has to be the new SM, as there may be numerous other alternatives, and if some of these signals go away, we may even do with a more minimal model.
Section \ref{sec:moretheory} discusses some of the ramifications of this model, like searches for singly- and 
doubly-charged scalars, as well as its further tests at the LHC and future leptonic colliders. 
Section \ref{sec:conclude} summarises and concludes the paper. Some details have been relegated to the Appendix.

\section{The Indications and Consequences}  \label{sec:signals}
We will assume no CP violation in the scalar sector, so that all the new resonances 
are CP-eigenstates. 
The CP-even states, as may be surmised from the decay pattern, will be denoted either by $h$ or $H$, 
the CP-odd resonances by $A$, and the charged scalars by $H^\pm$ and $H^{\pm\pm}$, with
subscripts denoting their (tentative) masses, {\em i.e.}, the approximate position of the peak of the 
resonance, in GeV.
 One may note that the heavier resonances, like $H_{650}$, 
are quite broad and the mass value
is the approximate position of the peak, which is prone to a significant uncertainty with such preliminary data. 
In this notation, as already mentioned,
the SM-like Higgs boson will be called $h_{125}$. 

A resonance may be observed in different channels, with $p$-values $p_1$, $p_2$, etc, where $p$ denotes the 
probability of observing the event assuming the null hypothesis. 
The $p$-values are easily obtained from the $\sigma$-values denoting the significances, as quoted by the 
individual experiments. The combined $p$-value is obtained following the method outlined in \cref{app:significances}. 
We will use the numbers quoted by the experiments only, and refrain from using any data for which such numbers are
not explicitly quoted, and then combine them following the prescription mentioned
in \cref{app:significances}.
Also, we will be dealing with only global significances, not the local ones, so that the LEE is taken care of.

\begin{table}[h]
\begin{center}
\begin{tabular}{||c||c||c|c|c||c||}
\hline
New scalar & Process studied & Local & Global & Combined & Reference \\
      &    & Significance & Significance & Significance & \\
\hline
 $h_{95}$     & $\to \gamma\gamma$ & $2.9\, \sigma$ & $1.3\, \sigma$ &            & \cite{CMS:2018cyk,CMS:2015ocq,Biekotter:2022jyr}\\
                    & $\to \tau^+\tau^-$  & $2.6$--$3.1\, \sigma$& $2.3$--$2.7\, \sigma$ &     $2.4$--$2.75\, \sigma$          &   \cite{CMS:2022rbd} \\
                    & $Z^*\to Zh_{95}\to Zb\bar{b}$ & $2.3\, \sigma$    & not quoted  &  $3.1$--$3.4\, \sigma$   &       
                    \cite{LEPWorkingGroupforHiggsbosonsearches:2003ing}\\
\hline
$H_{650}$   & VBF\,, $\to W^+W^-$  & $3.8\, \sigma$ &  $(2.6\pm 0.2)\, \sigma$  &            &  \cite{CMS:2022bcb} \\
                   & $\to ZZ$  &  $2.4\, \sigma$   &  $0.9\, \sigma$  &  $\left(4.08^{+0.12}_{-0.11}\right)\, \sigma$        &   \cite{ATLAS:2020tlo,ATLAS:2021kog} \\
                   & $\to h_{95}h_{125}$  &  $3.8\, \sigma$      & $2.8\, \sigma$  &       &     \cite{CMS:2022tgk}        \\
                   & $\to A_{400}Z \to \ell^+\ell^-t\bar{t}$ & $2.85\sigma$ & $2.35\sigma$ &  & \cite{ATLAS:2023zkt}\\
                   \hline
$A_{400}$   & $\to t\bar{t}$   &    $3.5\, \sigma$    & $1.9\, \sigma$ &     $3.17\, \sigma$    & \cite{CMS:2019pzc}    \\
                   & $\to ZH_{320}\to Zh_{125}h_{125}$ & $3.8\, \sigma$ & $2.8\, \sigma$   &   & \cite{ATLAS:2022wti}  \\
\hline
\end{tabular}
\caption{The existing indications for some scalar resonances, apart from $h_{125}$.
}
\label{tab:significance}
\end{center}
\end{table}
A summary of the signals is shown in Table \ref{tab:significance}. We would like to warn our readers again that the 
resonance peak positions are not at all precise, and for broad resonances, we club all the signals 
together that lie within $\sim 10\%$ of the tentative peak. We also do not consider the case of degenerate 
or close-lying resonances just for simplicity, although this is a distinct possibility for broad resonances. 
We have made a conservative estimate of the 
global significance of each resonance, and have included only those where this is above $3\sigma$. The case of $h_{95}$ and $H_{650}$
merits further discussion, for which we refer the reader to
Subsections \ref{sub:h95} and \ref{sub:h650} respectively.

\subsection{$h_{95}$} \label{sub:h95}
The CMS Collaboration \cite{CMS:2018cyk}, from the first year of the Run 2 data,
 found the hint of a resonance at about 95 GeV, decaying into two 
photons. A similar signal with slightly less significance was also observed in the Run I data \cite{CMS:2015ocq}.
Combined, this gives a local significance of $2.9\sigma$  \cite{Biekotter:2022jyr, SGS}, 
and a modest global excess of $1.3\sigma$.
An analysis for light resonances decaying into two photons has also been released by CMS 
\cite{CMS:2023yay}. 

This resonance has also been observed to decay to $\tau^+\tau^-$ with a local (global) significance of $2.6\, (2.3) \sigma$ 
\cite{CMS:2022rbd,CMS:2022goy}. 
One may note that these numbers are obtained with the assumption that the parent scalar resonance peaks at 95
GeV; CMS actually found it at 100 GeV, with a local (global) significance of $3.1\, (2.7) \sigma$ \cite{CMS:2022goy}\footnote{The authors of Ref.~\cite{2205.03187} showed that the cross-section
for $pp\to t\bar{t} h_{95}$, $h_{95}\to \tau^+\tau^-$ may be even larger than the corresponding 
cross-section with $h_{125}$. However, one may 
note that the background coming from $t\bar{t}Z$, $Z\to \tau^+\tau^-$ is 
large, and this decay channel for a scalar whose mass is so close to 
$M_Z$ may still remain hidden at the LHC ({\em e.g.}, a Z-veto may severely suppress the 
signal). Of course, if the backgrounds can be controlled, this should be a very 
promising channel, maybe in upcoming leptonic colliders.}.
Thus, we have the following from LHC:
\begin{align}
&\mu_{\gamma\gamma} = \frac{\sigma(pp \to h_{95} \to \gamma\gamma)}{\sigma(pp\to \phi\to
\gamma\gamma)} = 0.33^{+0.19}_{-0.12}\,, \label{eq:h95gamgam}\\
&\mu_{\tau^+\tau^-} = \frac{\sigma(pp \to h_{95} \to\tau^+\tau^-)}{\sigma(pp\to \phi \to
\tau^+\tau^-)} = 1.2\pm 0.5\,, \label{eq:h95tautau}
\end{align}
where $\phi$ is a fictitious 95 GeV Higgs boson, {\em i.e.}, a spin-0 object of mass 95 GeV whose couplings to 
the gauge bosons and fermions are exactly like the SM Higgs. 
The local significances are all less than $3\sigma$; a combined global significance reaches $2.75 \, (2.4) \sigma$ 
if one uses the higher (lower) significance value from Ref.\ \cite{CMS:2022rbd,CMS:2022goy}, 
as can be seen in Table \ref{tab:significance}. 

Earlier, LEP also found a similar excess close by \cite{LEPWorkingGroupforHiggsbosonsearches:2003ing}, 
at about 98 GeV, with a local significance of $2.3\sigma$, where we need to take into account 
the possibility of coarser mass resolution so that these two hints may come from one and the same particle:
\begin{equation}
\mu_{b\bar{b}} = \frac{\sigma(e^+e^- \to Zh_{95} \to Zb\bar{b})}{\sigma(e^+e^-\to Z\phi \to Zb\bar{b})} =
0.117\pm 0.057\,. \label{eq:h95LEP}
\end{equation}
Thus, if the branching fraction of $h_{95}\to b\bar{b}$ is the same as that of a SM-like 
95 GeV scalar, the $ZZh_{95}$
coupling is approximately $0.33\pm 0.09$ times the $ZZ\phi$ coupling\footnote{The significance quoted in the LEP paper
is a local one, no global significance has been quoted. If we combine the local significance of LEP with the global 
numbers of CMS, the combined significance just goes above $3\sigma$. This is in contrast with Ref.\ 
\cite{Biekotter:2022jyr} that claimed a combined significance of more than $4\sigma$, which can be obtained by 
combining local significances only and neglecting the LEE.}. 
This shows that, unless the $ZZh_{95}$ coupling is much smaller than the $ZZ\phi$ one, 
(i) $h_{95}$
must have a significant SU(2) doublet admixture, and (ii) a single SU(2) scalar doublet giving mass to all the fermions
is not enough to explain the decay pattern, as it would have led to similar values for $\mu_{\tau^+\tau^-}$ 
and $\mu_{b\bar{b}}$. The LEP indication in $b\bar{b}$ is however prone to controversy, as stressed in  \cite{Janot:2024ryq}.

\subsection{$H_{650}$}   \label{sub:h650}

It was first pointed out in Ref.\ \cite{Cea:2018tmm} that an analysis of the then total LHC data of 113.5 fb$^{-1}$
gives the signal of a broad resonance, of width ${\cal O}(100)$ GeV, that peaks between 600 and 700 GeV, with total
number of signal events given by
\begin{equation}
N_{\rm sig} = 34.21^{+12.78}_{-6.03}\,.
\end{equation}
We will rather work with the more recent updates published by the LHC Collaborations themselves.
As the properties of $H_{650}$ constitute an important part of this paper, we will briefly summarise the experimental 
results.

The resonance $H_{650}$ has been indicated in various channels. The most important results came from CMS
\cite{CMS:2022bcb}, who found the cross-section of $pp\to H_{650}\to W^+W^-$ to be 160~fb, assuming the 
production to be purely through the VBF process. This has a local (global) significance of $3.8\, (2.6)\sigma$. The 
significance becomes insignificant if the production, at least a large part of it, goes through ggF.

The ATLAS Collaboration  \cite{ATLAS:2020tlo,ATLAS:2021kog} looked for any possible resonance in the $ZZ$
decay channel in their 139 fb$^{-1}$ of data, 
ultimately leading to the 4-lepton as well as $\ell\nu\ell\nu$ signals. They found a mild excess of 
local (global) significance
$2.4 \, (0.9)\sigma$ \cite{ATLAS:2020tlo}, peaking at 620 GeV. Considering the large width of this resonance, 
we may assume that this is the identical resonance that we call $H_{650}$. 
An important point to note is that for ATLAS too, the significance,
however mild, shows only if the production is assumed to be entirely through VBF. This, of course, does not mean 
that ggF production is absent; it is just that such events are swamped under the background because of their 
topology. Also, the broadness of the line shape means that there is always the possibility of the signal being a 
superposition of more than one closely spaced resonances. While this is an interesting alternative, we will stick
to the simpler assumption of only a single resonance in the present paper. 
One may, however, note that the null result of CMS in the 4-lepton channel is consistent with a small enough
$ZZH_{650}$ coupling and/or a broad resonance \cite{CMS-ZZ}.

CMS \cite{CMS:2022tgk} also found a $3.8 \, (2.8)\sigma$ local (global) excess at 650 GeV for the decay channel 
$H_{650}\to h_{95}h_{125}$. The exact analysis of this channel depends on the scalar potential.

Finally, let us mention the ATLAS result \cite{ATLAS:2023zkt}, which looked for the channel $A \to HZ \to t\bar{t}
\ell^+\ell^-$, where $H$ and $A$ are generic neutral CP-even and CP-odd scalars. There is an excess of local
(global) significance $2.85\, (2.35) \sigma$ for $(m_A,m_H) = (650,450)$ GeV. However, one may note that apart from
being opposite CP neutral scalars, the CP assignment of the parent and the daughter is quite arbitrary; it could
very well be $(m_H,m_A) = (650,450)$ GeV. This is what we will assume in this paper. 

We convert all the significances to corresponding $p$-values and then combine them following the method outlined 
in \cref{app:significances}. 
Taking everything together, the global significance comes out to be just above $4\sigma$. 
This, therefore, is the strongest evidence of a new resonance, and we will therefore mostly focus upon this, 
as well as $h_{95}$, while the other resonances play a secondary role.

We examine now in some detail how to extract the  $H_{650}W^+W^-$ coupling from the CMS data of the production 
cross-section $\sigma(p p \to H_{650} \to W^+W^-) $. {The excess reported by CMS \cite{CMS:2022bcb} for $ \sigma(p p \to W^+ W^- \to 2\ell2\nu)$ is found to be consistent with a heavy SM-like scalar, where the data and the SM-like Higgs expectation intersect around a mass of 650~GeV and a $\sigma(p p \to H_{650} \to W^+W^-) $ cross-section around 160~fb, under the assumption that the latter  proceeds only through VBF. For a SM-like Higgs of mass 650 GeV, 
the theoretical VBF production cross-section $\sigma^{(\rm SM)}_{\rm VBF}$ is around 272~fb at $\sqrt{s}=13$~TeV \cite{CERNreport}. Thus, assuming no ggF contribution and neglecting consistently the $H_{650}$ partial width to fermions, this leads to the estimate  
$\sigma_{p p \to H_{650} \to W^+W^-}\simeq \sigma^{(\rm SM)}_{\rm VBF}\times {\rm BR}^{*(\rm SM)}_{H_{650} \to W^+W^-} \simeq 184~{\rm fb}$, where the star indicates the neglect of fermion contributions to the total $H_{650}$ width, and we used the SM-like updated widths 
\cite{CERNreport}.\footnote{
\label{footnote:3}
We used here a narrow-width approximation (NWA) which is not fully justified, since
$\Gamma/M \sim 18\%$. However, this allows for a consistent comparison with \cite{CMS:2022bcb} where the same approximation has been used to obtain the red line of their Figure 4.}
\begin{align}
\Gamma_{H_{650} \to WW} \simeq 81~\rm{GeV}\,,\ \ \ \Gamma_{H_{650} \to ZZ} \simeq 39~\rm{GeV}\,.
\end{align}
This theoretical cross-section, being in excess of the 160~fb reported by CMS under the same VBF dominance assumption, can be viewed as reflecting the effect of the experimental selection of the VBF-like category in the CMS analysis, that we estimate to be around 13\%. 
To access the $H_{650}W^+W^-$ coupling, one needs to separate the WW-fusion (WBF) from the ZZ-fusion (ZBF) contributions to the VBF cross-section. This is not straightforward, if meaningful at all, due to interference terms among s-, t- and u-channels in $q q \to q q H$ \cite{Dicus:1985zg}. However, the latter are expected to be small for a SM-like Higgs, especially when VBF cuts selecting hard forward or backward jets in $pp \to H j j$ are applied to reduce ggF contributions \cite{Ciccolini:2007ec,Ciccolini:2007jr}. Assuming this separation is obtained from a theoretical simulation, one can then in principle write a correlation between the (reduced) couplings $H_{650}W^+W^-$ and $H_{650}ZZ$ by requiring $\sigma_{\rm VBF}\times {\rm BR^*}_{H_{650} \to W^+W^-} \simeq 160$~fb. However, this will be consistent only if the separation between the ZBF and WBF contributions to the VBF is performed within the CMS categories selection procedure which is not available to us. Fortunately, one can proceed differently, avoiding the use of the absolute magnitude of the cross-section given by CMS. This is obtained by using the fact, noted above, that at 650~GeV the data cross-section coincides with the SM-like expectation. One can thus simply write
\begin{align}
 \sigma_{\rm VBF}\times {\rm BR}^*_{H_{650} \to W^+W^-}= c  \, \sigma^{(\rm SM)}_{\rm VBF}\times {\rm BR}^{*(\rm SM)}_{H_{650} \to W^+W^-}. \label{eq:xsec-data-SM}
\end{align}
The left-hand side depends on the reduced couplings
$\kappa_W^{H_{650}}$ and $\kappa_Z^{H_{650}}$ as well as on
the SM-like separate contributions to the VBF cross-section with appropriate cuts, 
$\sigma_{\rm WBF}^{\rm SM}$ and $\sigma_{\rm ZBF}^{\rm SM}$, while the right-hand side is now expressed formally in terms of the latter cross-sections rather than taking it as a number. In this way no explicit assumption is needed about category selection criteria or VBF cuts, except that it should be the same on both sides of \cref{eq:xsec-data-SM}).
The use of narrow width approximation (NWA) 
in \cref{eq:xsec-data-SM} is safe since the ensuing error is expected to be roughly the same on both sides of the equation, see also \cref{footnote:3}. We have, however, introduced a multiplicative factor $c$ to account for possible order one corrections. 
The ensuing correlation between $\kappa_W^{H_{650}}$ and $\kappa_Z^{H_{650}}$ takes the form: 
\begin{equation}
\left|\kappa_Z^{H_{650}}\right|^2= \frac{\left(\left|\kappa_W^{H_{650}}
\right|^2 - c  \, (1 + r) {\rm BR}^{*(\rm SM)}_{H_{650} \to W^+W^-} \right)\left|\kappa_W^{H_{650}}\right|^2
- c  \, (1 + r) {\rm BR}^{*(\rm SM)}_{H_{650} \to t \bar{t}} \left|\kappa_t^{H_{650}}
\right|^2 }{ c  \, (1 + r) {\rm BR}^{*(\rm SM)}_{H_{650} \to ZZ} - r \left|\kappa_W^{H_{650}}\right|^2 }\,,
\ \ \ r=\frac{\sigma_{\rm ZBF}^{\rm SM}}{\sigma_{\rm WBF}^{\rm SM}}
\label{eq:kappa650corr}
\end{equation}
where we have allowed for later use a non-vanishing reduced coupling to the top-quark, $\kappa_t^{H_{650}}$, but 
the starred BR still indicates that it is evaluated without including the corresponding width  in the $H_{650}$ {\sl total} width. The ratio $r$ is not expected to depend 
on the applied categorisation method to separate ggF from VBF cross-sections. We estimate it to be around 
$r\simeq 0.37$\footnote{\label{footnote:6}This estimate has been obtained by generating the process 
$p p \to H_{650} + 2 \rm jets$ at $\sqrt{s}=13$~TeV with MadGraph \cite{Alwall:2014hca,Frederix:2018nkq} version {\tt MG5\_aMC\_v3.5.4}, with no loop-induced ggF contribution, applying VBF cuts and comparing the cross-section when the Z-boson is excluded in the process, to that when $\rm W^+, W^-$ are excluded. As a consistency cross-check of the reduction of interference terms, the sum of the latter two is found to agree very well with the VBF cross-section when the VBF cuts are applied.}.
From the structure of \cref{eq:kappa650corr} one sees that 
$\left|\kappa_W^{H_{650}}\right|$ has to remain in a finite range in order to be consistent with the 
positivity of $\left|\kappa_Z^{H_{650}}\right|^2$.  Plugging in the actual values of the starred BRs 
and $r$ for a $650$~GeV SM-like Higgs, the allowed range for $\left|\kappa_W^{H_{650}}\right|$ is found 
to be very narrow, the minimum being around $0.96\sqrt{c}$, 
and the maximum well below $1.1\sqrt{c}$ in order for $\left|\kappa_Z^{H_{650}}\right|$ to remain perturbative, say $\lesssim {\cal O}(1)$. 
We infer the following bounds:  
\begin{equation}
0.96 \sqrt{c} \, gM_W  + {\cal O}\left(\left|\kappa_t^{H_{650}}
\right|^2\right)\lesssim \left| g_{WWH_{650}} \right| \lesssim  (0.05 + 0.95 \sqrt{c})\,gM_W  + {\cal O}\left(\left|\kappa_t^{H_{650}}
\right|^2\right) , \label{eq:WWH650min}
\end{equation}
with $c$ of order one. 
The lower bound estimate coincides with neglecting the VBF production through $ZZ$ fusion, 
as well as the coupling to fermions. While we can guess that $ZZ$ fusion is less effective than $WW$ fusion from the ATLAS data, it is not completely negligible. Taking it into account
will increase further the coupling $g_{WWH_{650}}$ above the lower bound given in \cref{eq:WWH650min}. Moreover, allowing for a non-vanishing coupling to fermions, which has to remain small to be consistent with the CMS indications, would slightly increase the lower and upper bounds themselves. One may also note 
that the allowed range can be scaled up or down by roughly a factor $\sqrt{c }$.}
Finally, let us mention that any other $H_{650}$ decay channel beyond the SM-like ones can be accounted for by shifting  ${\rm BR^*}^{(\rm SM)}_{H_{650} \to t \bar{t}} \left|\kappa_t^{H_{650}}\right|^2$ in \cref{eq:kappa650corr} by the corresponding branching fraction,
\begin{align}
 {\rm BR^*}^{(\rm SM)}_{H_{650} \to t \bar{t}} \left|\kappa_t^{H_{650}}\right|^2 \rightarrow {\rm BR^*}^{(\rm SM)}_{H_{650} \to t \bar{t}} \left|\kappa_t^{H_{650}}\right|^2 + {\rm BR^*}_{H_{650} \to h_{1} h_{2}}  
 \label{eq:shift}
\end{align}
This would always decrease the value of $\left|\kappa_Z^{H_{650}}\right|$ as long as the denominator in \cref{eq:kappa650corr} remains positive. We will come back to this point in \cref{sec:results} when discussing the possible decays of
$H_{650}$ to the lighter scalars of the model.

\subsection{$A_{400}$, and $H_{320}$}  \label{sub:a400}
The best evidence of $A_{400}$ comes from the CMS analysis
(see Fig.\ 4 of Ref.\ \cite{CMS:2019pzc}) where the invariant mass
reconstruction of the final $t\bar{t}$ state was used. The 
analysis is extremely challenging, given the large QCD background which interferes with the signal. 
CMS has performed a spin-parity analysis, which apparently selects a CP-odd candidate at about 400 
GeV, with $\Gamma/M \sim 4\%$. This has a local (global) significance of $3.5\, (1.9) \sigma$. 

While this is a modest evidence, ATLAS has also observed $A_{400} \to Zh_{125}h_{125}$ with about $3.8\sigma$ local 
and $2.8\sigma$ global significance \cite{ATLAS:2022wti}. 
The final state Higgs bosons apparently come from another scalar resonance, $H_{320}$. 
Admittedly, this is slightly below our $3\sigma$ global 
criterion, but this channel does not suffer from combinatorial background, since it originates from $A_{400}$. 
Assuming CP conservation in the scalar sector, this makes $H_{320}$ CP-even, and hence $A_{400}$ CP-odd. 

We may also note that the daughter scalar appearing in the final decay channel of $H_{650}$, namely, $H_{650}\to
AZ$, may indeed be $A_{400}$, albeit with a slightly less significance. However, it still remains a good possibility that
the two resonances at 400 GeV and 450 GeV are one and the same.

\subsection{$h_{151}$, or $A_{151}$}   \label{sub:h150}
A combined analysis of both CMS and ATLAS data was performed by the authors of 
Ref.\ \cite{Crivellin:2021ubm}, where they
found a local (global) significance of $4.3 (3.9) \sigma$ for a neutral scalar at around 
151 GeV, with the particle decaying to $\gamma\gamma$ or $Z\gamma$.
This claim has been disputed in Ref.\ \cite{2109.13426}, with a recalculated significance of $4.1\, (3.5) \sigma$. 
In spite of its potential interest, we will not discuss this resonance any further in this paper,
since its hint is not explicitly mentioned in any official publication, including public notes, from either ATLAS or CMS.  
One may note that its CP-properties are still uncertain, so it can be either $h_{151}$ (CP-even) or $A_{151}$ (CP-odd). However, if it exists, it can easily be accommodated in our model, 
namely, 2HDeGM.

\subsection{On sum rules}    \label{ss:sumrules}
If the underlying theory is renormalisable, the scattering unitarity bounds give strong constraints on the couplings, 
in the form of sum rules. Such sum rules exist in the literature for a long time; we refer the reader
to, {\em e.g.}, Ref.\ \cite{Gunion:1990kf}. We will be interested in the sum rules involving gauge bosons and scalars only. 

Suppose we parametrise the coupling between two gauge bosons and one scalar in the following way:
\begin{equation}
V_{a\alpha} V_{b\beta} \phi_i \qquad : \qquad i g_{abi} \, g_{\alpha\beta} \,,
\end{equation}
where $\alpha$ and $\beta$ are the Lorentz indices. This leads to the following sum rules for 
$W^+W^- \to W^+W^-$ and $W^+W^-\to ZZ$ scattering processes \cite{Gunion:1990kf}, assuming the 
custodial symmetry to be valid\footnote{If custodial symmetry is not valid, the left-hand side of the second equation 
of (\ref{eq:sum}) is replaced by $g^2 M_Z^4 \cos^2\theta_W/M_W^2$.} so that $\cos^2\theta_W = M_W^2/M_Z^2$:
\begin{eqnarray}
WW\to WW &:& g^2 M_W^2 = \sum_k g^2_{W^+W^-\phi_k^0} - \sum_\ell g^2_{W^+W^+\phi_\ell^{--}}\,,
\nonumber\\
WW\to ZZ &:& g^2 M_Z^2 = \sum_k g_{W^+W^-\phi_k^0}g_{ZZ\phi_k^0} - \sum_\ell g^2_{W^+Z\phi_\ell^-}\,,
\label{eq:sum}
\end{eqnarray}
where the summations are over all scalar states. 
If CP is not conserved in the scalar sector, the sum rules also depend on the CP-phases of the scalars.
 
One may draw the following conclusions:
\begin{itemize}
\item If the couplings of $h_{125}$ to $W^+W^-$ and $ZZ$ are exactly like the SM (which we call the alignment
limit), then no other neutral scalar can couple to $W^+W^-$ or $ZZ$ unless doubly charged scalars exist. This is true 
in the exact alignment limit, but even if there is a small deviation, the coupling of other neutral scalars to $W^+W^-$
or $ZZ$ would be tiny.
\item If doubly charged scalars exist, there must be some charged Higgs that couples to $W^+Z$. The reverse is also
true \cite{Gunion:1990kf}. 
\item 
The first equation of (\ref{eq:sum}) can be rephrased as
\begin{equation}
 \left[g^{(\rm SM)}_{WWh}\right]^2 = \sum_i g_{W^+ W^- \phi^0_i}^2 
- \sum_k \left\vert g_{W^- W^- {\phi_k}^{++}}\right\vert^2,  \label{eq:sumrule}
\end{equation}
where $g^{(\rm SM)}_{WWh}$ is the SM $WWh$ coupling. In writing Eq.\ (\ref{eq:sumrule}) one assumes 
CP conservation and all the neutral scalars $\phi^0_i$ to be CP-even eigenstates, and for simplicity 
$\rho=1$ at the tree-level\footnote{The sum rule, which essentially means that unitarity in $2\to 2$ 
scattering must not be violated, comes from longitudinal $W^+W^+\to W^+W^+$ scattering. The neutral scalars 
lead to $s$-channel processes, while a doubly charged Higgs leads to a $t$-channel process. This new contribution 
softens the sum rule constraint.}. 
\end{itemize}
Once we have a renormalisable model to start with, the sum rules are automatically satisfied. However, a more
challenging and interesting way is to determine the couplings experimentally and see how far the sum rules are 
satisfied; this may lead to the discovery of new scalars.

\subsection{Inadequacy of Higgs doublets, and the canonical Georgi-Machacek model}    \label{ss:fail}
So far, all the new resonances that we have talked about are neutral scalars. There has been no definite hint
of any charged scalar, but there are strong constraints, both direct and indirect. One may ask why the spectrum 
--- with the usual {\em caveat emptor} that some, most, or all of them may vanish in near future --- cannot be 
accommodated in a multi-Higgs doublet model, or even in a more exotic extension like the canonical
GM model. 

To answer this, let us just focus on $H_{650}$, the exact mass value being of little relevance right now.  
If we take the data at face value and assume that it will stand the test of time, 
the analysis of Subsection \ref{sub:h650} 
leads to the following:
\begin{itemize}
\item
The scalar sector of the SM has to be extended. The 125 GeV resonance,
$h_{125}$, behaves almost exactly like the Standard Model Higgs boson, $h_{\rm SM}$. Thus, to a first approximation, 
we may assume that the extended scalar sector is at the `alignment limit', {\em i.e.}, the gauge and fermionic
couplings of $h_{125}$ are almost exactly that of $h_{\rm SM}$. There may be small deviations, which are generally 
parametrised by the $\kappa$-parameters. We will keep $|\kappa-1| < 0.1$, 
consistent with Eq.\ (\ref{eq:CMS-SMkappas}),  a conservative estimate for all couplings of
$h_{125}$, which means a maximum deviation of 10\% from the corresponding SM couplings. 

\item
\Cref{eq:WWH650min} shows that $\left|g_{WWH_{650}}\right|$ is confined to a narrow range with large values.
The implication of such a large $\left|g_{WWH_{650}}\right|$ is a clear violation of the
sum rule, Eq.\  (\ref{eq:sumrule}), as the sum on $i$ includes now at least $h_{125}$ and $H_{650}$ and the SM-like $WWh_{125}$ coupling almost saturates the sum rule, 
{\sl unless there are doubly charged scalars in the spectrum.} 
In order to fulfil the sum rule within a coherent extension of the SM, let us examine the various possibilities. 
Considering only $\rm SU(2)_L$ singlets would obviously not work. On the one hand, a
doubly-charged state would then not couple to $W^+W^+$, and thus will not help in restoring the sum rule, and
on the other hand, $H_{650}$
would also then dominantly be a gauge singlet and its coupling to $W^+W^-$ would contradict
the CMS indications.
Adding just an $\rm SU(2)_L$ doublet to the SM doublet would not work either. Indeed this would 
require both the components to be electrically charged to account for a doubly-charged state. This 
second doublet should then not develop a VEV, to avoid spontaneous breaking the electric charge 
conservation. However, even then, one would be left with 5 physical degrees of freedom, four of which 
are taken by the charged states and no room is left for a second neutral state. Adding a third doublet with a neutral and a charged component, there are now 9 physical states, of which only two would-be 
CP-even neutral scalars, thus discarding the inclusion of $h_{95}$ in the spectrum. 
{Going further with more singlets or doublets, although a logical possibility, would be at the expense of minimality. Thus, avoiding the unusual construct of a doublet with two charged components,} one may say that multi-Higgs doublet models fail to explain the spectrum and its characteristics. Thus, there must be SU(2) representations higher than doublet.

\item
The minimal way of including a doubly-charged scalar respecting the gauge symmetries of the model would be by adding an $\rm SU(2)_L$ complex triplet.  A major problem with scalar multiplets higher than doublet is that they generally do not respect the custodial symmetry (CS), which keeps $\rho=1$ at the tree-level. For instance, adding only one complex triplet like in the type-II seesaw model, and enforcing $\rho$ to remain consistent with unity, Eq.\ (\ref{eq:rho-expt}), induces a totally different phenomenology from the one expected for $H_{650}$ or its (doubly) charged partners, suppressing 
single productions at the LHC (see, {\em e.g.}, Ref.\ \cite{Ducu:2024xxf} for a very recent appraisal.) Thus one complex
 triplet would not work in our case. One elegant way out that ensures $\rho=1$ without further constraints at the 
 tree-level, was suggested by Georgi and Machacek
(GM) \cite{Georgi:1985nv}. We refer the reader to Ref.\ \cite{KMP22} for a detailed discussion on CS for the GM models. 
Thus, the canonical GM model could be a good starting point. One may even look at the extended GM proposed in
Ref.\ \cite{KMP22}. The GM model has one doublet, one real triplet, and one complex triplet of scalars under the weak
SU(2). After spontaneous symmetry breaking, the custodial symmetry still remains intact, and the physical scalars
can be grouped in custodial multiplets, for which there are two singlets, one triplet, and one 5-plet. 
\item
Does the canonical GM model have enough scalars to fit the bill? Assuming that 
$H_{650}$ is a CP-even neutral scalar, there are only three such
objects in the GM model. Can we ascribe them to $h_{95}$, $h_{125}$, and $H_{650}$? The answer is, unfortunately, no.
The reason is twofold. The first and major objection is that in the GM model, the $H_{650}ZZ$ coupling is
$2/\cos^2\theta_W \approx 2.6$ times as strong as the $H_{650}WW$ coupling \cite{HKL14}, if $H_{650}$ is a custodial singlet or member of the custodial 5-plet, respectively.
In both cases this is clearly contradictory to the hints. A milder objection is that in the GM model, the custodial 5-plet consists 
entirely of SU(2) triplets and hence the neutral member does not couple to fermions. Thus, there 
would be no ggF production of $H_{650}$\footnote{$pp \to H_{650}\to WW$
through ggF has not yet been detected, which is probably due to the nature of the background.}. 
\end{itemize}

\section{The 2-Higgs Doublet extended Georgi-Machacek model} \label{sec:theory}
We would like to propose a new model here, which we call Two-Higgs Doublet 
extended Georgi-Machacek Model (2HDeGM for short). 
The extension from the canonical GM model occurs in two steps. First, we add one more 
$\rm SU(2)_L$ scalar doublet with the same weak hypercharge as the first doublet. Second, we
impose CS only on the gauge part of the Lagrangian and not on the scalar
potential, in the same vein as Ref.\ \cite{KMP22}. The latter condition removes the mass degeneracy
among the scalar states in any nonsinglet custodial multiplet.
Thus, 2HDeGM consists of
two SU(2) doublets, $\Phi_1$ and $\Phi_2$
(with VEVs $v_1$ and $v_2$ respectively), a real triplet $\Xi$ and a complex triplet $X$
(both with VEV $u$), and with the SU(2)$_L$ and U(1)$_Y$ quantum numbers $(T,Y)$ as
\[
\Phi_1:\  (\textstyle\frac12,1)\,,\ \ \
\Phi_2:\ (\textstyle\frac12,1)\,,\ \ \
\Xi:\ (1,0)\,,\ \ \
X:\ (1,2)\,,
\]
with the electric charge $Q=T_3+Y/2$.

The CP-even neutral members of these multiplets are denoted by $\phi_1^0$, $\phi_2^0$, $\xi^0$, and $\chi^0$ respectively.
This is a minimal extension of the canonical GM model, and by construction preserves CS.
We do not show the full scalar potential, but 
we will assume CS to be respected in the gauge sector only, keeping $\rho=1$ at the tree-level.
It need not be respected by the scalar potential, as is explained in Ref.\ \cite{KMP22}.
For the subsequent analysis, we will assume that CS is not respected in the scalar potential.

There are at least four two-Higgs doublet models (2HDM) that forbid tree-level flavour-changing
neutral current through some discrete symmetries on the scalar sector. They are typically
called Type-I, Type-II, Type-X (or lepton-specific, sometimes called Type-III), and Type-Y (or flipped). We will follow
the notation of Ref.\ \cite{Branco:2011iw}, so that the couplings of the SU(2) scalar doublets
$\Phi_1$ and $\Phi_2$ with fermions look as follows:\\
1. Type-I: $\Phi_2$ couples with all quarks and charged leptons, $\Phi_1$ has no fermionic coupling.\\
2. Type-II: $\Phi_2$ couples with weak isospin projection $I_3=+\frac12$ fermions, and
$\Phi_1$ with $I_3=-\frac12$ fermions.\\
3. Type-X: $\Phi_2$ couples with quarks and $\Phi_1$ couples with leptons. Thus, Type-I and
Type-X have identical roles in quark flavour physics.\\
4. Type-Y: Couplings with the quarks are identical to that of Type-II, but $\Phi_2$ couples
with charged leptons.\\
Thus, as far as quark flavour physics is concerned, Type-I and Type-X behave identically,
and Type-II and Type-Y are also indistinguishable.

There are 17 scalar fields to start with, and 3 of them are eaten up by the gauge bosons. The construction obviously
respects CS, and the custodial multiplets have to be
\begin{equation}
{\bm 5} \oplus {\bm 3} \oplus {\bm 3} \oplus {\bm 3} \oplus {\bm 1} \oplus {\bm 1} \oplus {\bm 1}\,.
\end{equation}
That this is the unique decomposition into irreducible multiplets can be easily understood just by counting the 
number of charged fields.
Note that these {\em need not be} the mass eigenstates \cite{KMP22}.

One of the ${\bm 3}$s is the Goldstone triplet. In the final spectrum, we have 4 CP-even and 2 CP-odd neutral
scalars. We may identify the CP-even neutrals with $h_{95}$, $h_{125}$, $H_{320}$ and $H_{650}$ respectively. 
This is what we will assume in the rest of the paper, and proceed accordingly.
Similarly, we may identify the CP-odd neutrals with $A_{400}$ and something else, yet to be detected. For example, 
this can be $A_{151}$. 

There are 3 independent VEVs, and so 3 custodial singlets. The neutral members of the higher multiplets,
${\bm 5}$ or ${\bm 3}$, cannot develop a VEV, as that will break the CS. 
One of the CP-even neutral scalars must go into {\bf 5}; let us call that $H_5$.
It should be a linear combination of
$\phi_1^0\sin\beta - \phi_2^0\cos\beta$ (where $\tan\beta = v_2/v_1$), and $\chi^0-\sqrt{2}\xi^0$.
However, vertices like $H_5W^+W^-$ or $H_5 ZZ$ can exist \cite{HKL14}. 
One may note that $H_5$ need not be a mass eigenstate, except $H_5^{\pm\pm}$.

$h_{95}$ decays to $\tau^+\tau^-$ and $b\bar{b}$, so it must contain some doublet fields\footnote{There are three
bases we will be talking about: the gauge basis, with two doublets and two triplets, the custodial basis, whose 
irreducible representations are $\bm{5,3,3,3,1,1,1}$, and the mass basis.}. The 125 GeV resonance, $h_{125}$, 
must be very close to the SM Higgs boson $h_{\rm SM}$, {\em i.e.}, in the alignment limit. 
$H_{650}$ may or may not have any 
doublet components; whether it is produced in ggF is still not clear. 

The SU(2) triplet fields do not couple to fermions (except, possibly, neutrinos, which are outside our scope), but the
doublet fields do. To prevent tree-level flavour-changing neutral current, we consider two of the most popular
choices: Type I, where the doublet $\Phi_1$ has no fermionic coupling, and the doublet $\Phi_2$ gives mass to 
all the fermions, and Type II, where $\Phi_1$ ($\Phi_2$) gives mass to isospin projection $-\frac12$ ($+\frac12$) 
fermions. Other choices are also possible.

Let us also recall that on top of the unitarity sum rules considered in \cref{ss:sumrules}, one can 
also impose unitarity requirements on scattering processes involving only the scalar states. The latter 
lead to constraints involving combinations of the couplings in the scalar potential. These can translate 
into upper bounds on the physical scalar masses, as shown for example in Ref.\ \cite{AK08} for the 
GM model when a $Z_2$ symmetry is imposed. However, relaxing the latter discrete symmetry, 
one can identify decoupling limits where the mass upper bounds are totally relaxed \cite{HKL14}. 
In the present study we do not need to specify the scalar potential for the extended Higgs sector. 
We will assume  it to be the most generic compatible with the gauge symmetries and renormalisability, 
so that one expects to find decoupling limits despite the unitarity constraints,  and thus no upper bounds 
on the scalar masses. Note, however, that even imposing some extra global symmetries, the bounds 
of \cite{AK08} need not hold any longer in our case; there are more $|{\rm in}\rangle$ and 
$|{\rm out}\rangle$ states, and, unlike the GM model, 
the scalar masses are no longer expressible
in terms of the common triplet mass $m_3$ and common 5-plet mass $m_5$. 
The CMS bound on the mass of $H^{++}$, coming from $H^{++}\to W^+W^+$, need not be valid in the 
presence of competing channels opening up, like the decay to one charged scalar and $W^+$, or two 
singly charged scalars. This opens up the possibility of a comparatively large triplet VEV $u$ than what is usually 
found in the context of the GM model. 


\subsection{Convention} \label{sec:conv}
For any generic multiplet $\Phi$, the neutral component $\Phi_{Q=0}$ gets a VEV:
\begin{eqnarray}
  \Phi_{Q=0} = \begin{cases} v + {1 \over \surd2} (\Phi^0 + i {\Phi'}^0) 
    & \mbox{if  the multiplet $\Phi$  is complex,}    \\ 
  v +\Phi^0  & \mbox{if  the multiplet $\Phi$  is real},
  \end{cases}
  \label{vev-define}
\end{eqnarray}
In this convention, $v=174$ GeV in the SM, with $M_W^2 = \frac12 g^2 v^2$. 
The vacuum shifts for the CP-even neutral scalars of 2HDeGM are as follows:
$$
\phi_1^0 \to \phi_1^0+\sqrt{2} v_1\,, \ \ \phi_2^0 \to \phi_2^0 + \sqrt{2} v_2\,, \ \ 
\chi^0 \to \chi^0 + \sqrt{2} u\,,\ \ \xi^0 \to \xi^0 + u\,,
$$
and 
\begin{equation}
 M_W^2 = \frac12 g^2\left( v_1^2+v_2^2+4u^2\right)\, ,   
\end{equation}
where now 
\begin{equation}
   v=\sqrt{ v_1^2+v_2^2+4u^2}
   \label{eq:v} \, .
\end{equation}
We also define
\begin{equation}
u = \frac{v}{2}\, \sin\theta_H\,,\ \ \ v_1 = v\, \cos\theta_H\,\cos\beta\,,\ \ \ v_2 = v\, \cos\theta_H \, \sin\beta\,,
\end{equation}
so that 
\begin{equation}
\tan\theta_H = \frac{2u}{\sqrt{v_1^2+v_2^2}}\,,\ \ \ \tan\beta = \frac{v_2}{v_1}\,.
\end{equation}
While we take all the VEVs to be real, their relative signs are unconstrained, {\em i.e.}, $\beta$ and $\theta_H$
need not be fixed in the first quadrant. 
The custodial singlets of 2HDeGM are given by
\begin{equation}
S_1 = \phi_1^0\,,\ \ \ S_2 = \phi_2^0\,,\ \ \ S_3 = \frac{1}{\sqrt{3}} \left(\sqrt{2}\chi^0+\xi^0\right)\,,
\label{eq:cusmul}
\end{equation}
and the neutral component of the 5-plet, orthogonal to all of them, must be 
\begin{equation}
F_0= \frac{1}{\sqrt{3}} \left( \chi^0 - \sqrt{2}\xi^0\right)\,.
\end{equation}
The gauge-gauge-scalar terms in the Lagrangian, relevant for the sum rules, are \cite{KMP22}
\begin{equation}
{\cal L}_{\rm cubic} = \frac{g^2v_1}{2\sqrt{2}} S_1 \, \bm{W}\cdot \bm{W}
+ \frac{g^2v_2}{2\sqrt{2}} S_2 \, \bm{W}\cdot \bm{W} 
+ \frac{2g^2 u}{\sqrt{3}} S_3\, \bm{W}\cdot \bm{W} 
+ g^2 u \left[ \bm{W}\otimes \bm{W}\right]\cdot \bm{F}\,,
\end{equation}
where 
\begin{equation}
\bm{W}\cdot \bm{W} = 2W^+W^- + (W_3)^2\,,\ \ \ 
\bm{W}\otimes \bm{W}\big\vert_0 = \frac{1}{\sqrt{6}} \left[ -2W^+W^- + 2 (W_3)^2\right]\,.
\end{equation}
This immediately gives the couplings of the CP-even neutral scalars to $W^+W^-$.
To obtain the coupling with $ZZ$, one just has to replace $g^2 W_3^2$ with $(g^2/\cos^2\theta_W) Z^2$.
$\bm{F}$ is the 5-plet custodial multiplet, which is dotted with the 5-plet combination
$\bm{W}\otimes \bm{W}$. It is easy to check that the $W_3W_3 \xi^0$ terms cancel out, as they have to be. 

It is important to note here that ${\cal L}_{\rm cubic}$, by construction, respects the CS. As a consequence, 
if the potential is assumed to be that of the canonical GM which also respects
the CS, the physical $H_{650}$ mass eigenstate would be either one of the
custodial singlets or part of the custodial 5-plet. Its couplings to $WW$ and
$ZZ$ would then be uniquely fixed by
${\cal L}_{\rm cubic}$, bringing us back to the inconsistency related to these
couplings noted in the last item of \cref{ss:fail}. This justifies departing from the
canonical GM potential in the 2HDeGM model as previously stated.

Note that any combination of custodial singlets is a custodial singlet. For example, instead of $S_1$ and $S_2$, 
we can define
\begin{equation}
S_h = \phi_1^0\cos\beta + \phi_2^0\sin\beta\,,\ \ S_H = -\phi_1^0\sin\beta+\phi_2^0\cos\beta\,, 
\end{equation}
as two custodial singlets. Note that $\bra S_H\ket = 0$ and $\bra S_h\ket = 
\sqrt{v_1^2+v_2^2}$. 

So, the terms that we get from ${\cal L}_{\rm cubic}$ are
\begin{eqnarray}
{\cal L}_{\rm cubic} &=& g^2 \left[ 
\frac{v_1}{2\sqrt{2}}\, \phi_1^0 W_3^2 + \frac{v_1}{\sqrt{2}}\, \phi_1^0 W^+W^-  + 
\frac{v_2}{2\sqrt{2}} \, \phi_2^0 W_3^2 + \frac{v_2}{\sqrt{2}} \, \phi_2^0 W^+W^- \right.\nonumber\\
&& \left. +\sqrt{2}u\, \chi^0 W^+W^- + \sqrt{2} u\, \chi^0 W_3^2 + 2u\, \xi^0 W^+W^-\right]\,.
\label{eq:l-cubic}
\end{eqnarray}

Now, let us define the matrix ${\cal X}$ as the one that rotates the gauge basis to the mass basis:
\begin{equation}
\begin{pmatrix} h_{95} \\ h_{125} \\ H_{320} \\ H_{650} \end{pmatrix} = {\cal X}_{4\times 4} 
\begin{pmatrix} \phi_1^0 \\ \phi_2^0 \\ \chi^0 \\ \xi^0 \end{pmatrix}\,.
\label{eq:rotation}
\end{equation}
${\cal X}$ is orthogonal, so $H={\cal X}\Phi$ means $\Phi = {\cal X}^\top H$, or $\phi_i = x_{ji} H_j$, {where 
all $x_{ij}$s are real-valued satisfying,
\begin{equation}
 \sum_j x_{ij} x_{kj}  =\delta_{ik} \,.  \label{eq:unitarity}
\end{equation}} For example,
\begin{align}
\label{eq:h125_content}
h_{125} = x_{21}\phi_1^0+x_{22}\phi_2^0+x_{23}\chi^0+x_{24}\xi^0\,.
\end{align}

From Eq.\ (\ref{eq:l-cubic}), one gets 
\begin{equation}
\begin{aligned}
{\cal L}_{W^+W^-h} = \;\;  g^2\, W^+W^-\, &\left[
\left\{ \frac{v_1}{\sqrt{2}} x_{11} + \frac{v_2}{\sqrt{2}} x_{12} + \sqrt{2}u x_{13} + 2ux_{14} \right\}\, h_{95} \right.\\
& + 
\left\{ \frac{v_1}{\sqrt{2}} x_{21} + \frac{v_2}{\sqrt{2}} x_{22} + \sqrt{2}u x_{23} + 2ux_{24} \right\}\, h_{125} \\
& +
\left\{ \frac{v_1}{\sqrt{2}} x_{31} + \frac{v_2}{\sqrt{2}} x_{32} + \sqrt{2}u x_{33} + 2ux_{34} \right\}\, H_{320} \\
&+
\left. \left\{ \frac{v_1}{\sqrt{2}} x_{41} + \frac{v_2}{\sqrt{2}} x_{42} + \sqrt{2}u x_{43} + 2ux_{44} \right\}\, H_{650}\right]\,,
\label{eq:hww}
\end{aligned}
\end{equation}
where the obvious Lorentz indices have been suppressed.

Similarly,
\begin{equation}
\begin{aligned}
{\cal L}_{ZZh} =  \;\;  \frac{g^2}{\cos^2\theta_W}\, ZZ\, &\left[
\left\{ \frac{v_1}{2\sqrt{2}} x_{11} + \frac{v_2}{2\sqrt{2}} x_{12} + \sqrt{2}u x_{13} \right\}\, h_{95} \right.\\
& + 
\left\{ \frac{v_1}{2\sqrt{2}} x_{21} + \frac{v_2}{2\sqrt{2}} x_{22} + \sqrt{2}u x_{23} \right\}\, h_{125} \\
& +
\left\{ \frac{v_1}{2\sqrt{2}} x_{31} + \frac{v_2}{2\sqrt{2}} x_{32} + \sqrt{2}u x_{33}  \right\}\, H_{320} \\
&+
\left. \left\{ \frac{v_1}{2\sqrt{2}} x_{41} + \frac{v_2}{2\sqrt{2}} x_{42} + \sqrt{2}u x_{43}  \right\}\, H_{650}\right]\,,
\label{eq:hzz}
\end{aligned}
\end{equation}
Eqs.\ (\ref{eq:hww}) and (\ref{eq:hzz}) are consistent with the sum rules given in Ref.\ \cite{Gunion:1990kf}. For example, 
if all other $x_{ij}$s are zero except $x_{22}=1$, the sum on the r.h.s.\ of Eq.\ (4.1) of \cite{Gunion:1990kf} reduces to just one term, 
$g^4 v_2^2/2=g^2 M_W^2$. If one keeps all the terms and also brings in the doubly charged Higgs, that will give some
more nontrivial constraint on $x_{ij}$s. This is what we get from Eqs.\ (\ref{eq:hww}) and (\ref{eq:hzz}):
\begin{itemize}
\item 
$H_{650}$ need not be a pure 5-plet. Even if it is, the $x_{44}$ term creates a difference between $\sigma(pp\to
H_{650}\to WW)$ and $\sigma(pp\to H_{650}\to ZZ)$. If it is a pure 5-plet (this happens if the full 
scalar potential is also invariant under CS), then
\begin{equation}
\sqrt{v_1^2+v_2^2} \left( x_{41} \cos\beta + x_{42} \sin\beta\right) + u \left(x_{43} + \sqrt{2} x_{44}\right) = 0\,.
\end{equation}
\item
$h_{95}\to \tau^+\tau^-$ and $h_{95}\to b\bar{b}$ depend not only on $v_1$ and $v_2$ but also on $x_{11}$ 
and $x_{12}$, and definitely on what type of 2HDM we have in mind. 
\item 
Eq.\ (\ref{eq:sum}) gives the tightest constraint on this model. First, take the limit $v_1,v_2 \gg u$, so that we
can neglect $x_{i3}$ and $x_{i4}$ terms. This always makes $\sum_k {g^2_{WW\phi_k}} > g^2 M_W^2$. So we need
to bring in the doubly charged Higgs, but the $W^+W^+\chi^{--}$ coupling is just $g^2u$, and if $u$ is too small, it can never satisfy the unitarity sum rule Eq.\ (\ref{eq:sum}).
\end{itemize}

\subsection{Input  strategy} \label{sec:strategy}
{Note that \cref{eq:rotation} is in line with the implicit assumption of absence of explicit CP-violation in the potential, so that the} neutral scalar 
{mass matrix splits into two separate sectors of CP-even and CP-odd eigenstates. Requiring also real VEVs $v_1, v_2$ and $u$, we avoid discussing new sources of CP-violation and the possibly related constraints.}
To determine the elements of this orthogonal matrix, we need 6 inputs: 3 to determine the first row, 
2 for the second, and 1 for the third, and the rest get determined from the orthogonality conditions.  Unfortunately, it is
not straightforward to obtain these 6 parameters, and this is a more speculative job than just showing the violation 
of the sum rule. This is how we choose the 6 parameters:
\begin{itemize}
\item 
Three of them come from the coupling of $h_{125}$ with $t\bar{t}$, $W^+W^-$, and $ZZ$, consistent with 
Ref.\ \cite{CMSnature}. 
\item
Two of them are obtained from the couplings of $h_{95}$ to $t\bar{t}$ and $W^+W^-$. The decay $h_{95}\to
\gamma\gamma$ depends on both these couplings, both in production and in decay. We assume the ratio of ggF 
to VBF production for a 95 GeV SM Higgs-like scalar to be $90:10$, which is scaled by the corresponding coupling 
modifiers or the $\kappa$-factors. A similar thing happens for the decay too. This gives us allowed regions in the 
$\kappa_t$--$\kappa_W$ plane for $h_{95}$. If we assume the underlying 2HDM to be of Type-I, $\kappa_t=\kappa_\tau$, 
and we get another constraint on $\kappa_t$ from $h_{95}\to \tau^+\tau^-$. Of course, we need to check that 
the numbers are consistent with the LEP result on $e^+e^-\to Zh_{95} \to Zb\bar{b}$. How this is done will be discussed in Section \ref{sec:results}.
\item
Finally, the last number is obtained from the coupling of $H_{650}$ to $W^+W^-$. We do not consider the $H_{650}ZZ$ 
coupling as it is small and plagued with a large uncertainty. 
\end{itemize}

\subsection{Possible solutions and implications \label{sec:poss-sols}}
Based on \cref{eq:hww,eq:hzz}  and further assumptions on the Yukawa sector, we explore in this section various 
theoretical configurations in order to delineate the  $v_1,v_2,u$ and $x_{ij}$'s parameter-space regions that are 
consistent with the experimental indications described in \cref{sec:signals}. Obviously, one should start with the
best known state, namely $h_{125}$, to put constraints on the three VEV's and on its doublet and triplet content 
given by  $x_{2i}, (i=1,...4)$, cf. \cref{eq:h125_content}. As we will see this is quite constraining by itself due 
to the precise experimental knowledge of the $h_{125}$ couplings to gauge bosons and fermions. As shown in
 \cref{app:A1}, if the {\em reduced} couplings
of $h_{125}$ to a pair of $W$s and to a pair of $Z$s, denoted respectively $\kappa_W^{h_{125}}$ and $\kappa_Z^{h_{125}}$, 
are equal to each other, then, either $u=0$, or $x_{23}$ is correlated to $x_{24}$ through \cref{eq:x23x24_relation}. 
This holds true even if these reduced couplings differ from one, i.e. even if they are not exactly SM-like. 
The experimentally determined central values of the signal strengths or the fitted coupling modifiers 
\cite{CMSnature, toro}  indeed suggest that we are in such a configuration. 
This very constrained configuration can be somewhat relaxed within one or two standard deviations
of the measured quantities. In this case, combining \cref{eq:h125eq1,eq:h125eq2} leads to
\begin{align}
u \left(x_{2 3} - \sqrt{2} x_{2 4}\right)= \frac{v}{2} \delta 
\label{eq:ux_relation}
\end{align}
where $\delta = \kappa_Z^{h_{125}} - \kappa_W^{h_{125}}$. It is however instructive to treat first the case $\delta =0$ assuming \cref{eq:x23x24_relation}. Also the results strongly depend on what one assumes about
the Yukawa sector.

\subsubsection{Type-II Yukawa}
\label{sec:typeIIyukawa}
If the couplings of the $\Phi_1$ and $\Phi_2$ doublets to fermions are taken to be those of the Type-II 2HDM, then the reduced couplings of $h_{125}$ to the down- and up-type fermions read respectively
\begin{align}
    \kappa^{h_{125}}_d = \frac{v}{v_1} x_{21}\,, \ \ \kappa^{h_{125}}_u= \frac{v}{v_2} x_{22}\,.
    \label{eq:typeII_kappas}
\end{align}
and more generally
\begin{align}
    \kappa^{\mathcal{H}_a}_d = \frac{v}{v_1} x_{a1}\,, \ \ \kappa^{\mathcal{H}_a}_u= \frac{v}{v_2} x_{a2}, \ a=1,2,3,4,
\label{eq:typeII_kappas-gen}
\end{align}
where we define $\left( {\cal H}_a \right)_{a=1,2,3,4} = \left(h_{95}, h_{125}, H_{320}, H_{650} \right)$.
In particular
there is no extra freedom from the Yukawa couplings here;  the reduced couplings $\kappa^{\mathcal{H}_a}_d $ ($\kappa^{\mathcal{H}_a}_u$) 
being obtained by equating directly the SM expressions of the down-type (up-type) fermion masses to 
those of the Type-II.
The experimentally well-determined values of $\kappa^{h_{125}}_t$, $\kappa^{h_{125}}_b$, $\kappa^{h_{125}}_\tau$ and $\kappa^{h_{125}}_\mu$
will thus lead to further constraints
on the parameters of the scalar/gauge boson sector of our model. 

Since the experimental values of the $\kappa^{h_{125}}$'s are very close to one, let us illustrate first the
case $\kappa^{h_{125}}_t =\kappa^{h_{125}}_b=1$. Upon use of \cref{eq:typeII_kappas,eq:x23x24_relation,eq:h125eq3} in the normality condition $\sum_j x_{2j}^2   =1 $, one finds readily a simple relation between $u$ and $x_{24}$, 
namely $4 u^2= 3 x_{24}^2 v^2$.
Injecting this relation as well as \cref{eq:typeII_kappas} in \cref{eq:h125eq2} determines $u$ uniquely (up to a global sign):
\begin{align}
 u^2 = \frac34 \frac{1-\kappa_Z^{h_{125}}}{3  - \sgn(u) 2 \sqrt{6}} v^2   
\end{align}
This relation illustrates a tension in the Type-II case for obtaining a sufficiently large  value of $u$
necessary for the consistency of our approach, as explained in \cref{ss:sumrules}.
For instance, taking $\kappa_Z^{h_{125}} \simeq 1.04$ ($0.985$) 
and $v=174$~GeV, one finds
$u\simeq 22$~GeV ($\simeq-6.6$~GeV); these values are too small. Indeed, we anticipate here the needed value of $|u| \gtrsim 65$~GeV to ensure real-valued reduced couplings of  $h_{95}$, $H_{320}$ and $H_{650}$ to $t\bar t$, $W^+W^-$, and $ZZ$, the determination of which will be discussed in the following sections.  This calls for a relaxation of the two simplifying assumptions made so far, namely
taking $\delta \neq 0$ in \cref{eq:ux_relation} and/or   $\kappa^{h_{125}}_t \neq \kappa^{h_{125}}_b \neq 1$ within the allowed experimental error bars. In this case the analysis becomes much more involved and is actually included in a general strategy applicable to any of the Yukawa types. The reader is referred to
\cref{app:A1} for a description of its main ingredients. 

Suffice it to say here that a numerical scan, varying the reduced couplings within the experimental ranges given in 
\cref{eq:CMS-SMkappas}, still shows that the maximal $|u|$ does not exceed $\simeq 50$~GeV. 
However, the value can be raised further if $\kappa_b$ is negative. While right now the sign of $\kappa_b$ is a free
 parameter, there is a possibility that it may be determined unambiguously in future \cite{1403.4736}. 
 For our subsequent discussions, we will be rather conservative and treat both $\kappa_b^{h_{125}}$ and 
 $\kappa_\tau^{h_{125}}$ as positive numbers.
As pointed out above, such small values of $|u|$ fail to ensure real-valued couplings in the CP-even scalar sector.

\subsubsection{Type-X and Type-Y Yukawa}
\label{sec:typeXYyukawa}
The quarks in Type-X 2HDM get their masses either from $\Phi_1$ or from $\Phi_2$, exactly like Type-I 2HDM,
which is discussed below in detail. However, in Type-X the charged leptons get their masses from the other 
scalar like Type-II. The reduced couplings $\kappa_\tau^{h_{125}}$ and $\kappa_\mu^{h_{125}}$ being close to unity, 
the rationale remains the same as above
and the triplet VEV $u$ has a similar upper bound to what is found for Type-II. As for Type-Y 2HDM, the quarks get their masses exactly like in Type-II, so the triplet VEV $u$ has obviously the same upper bounds as what is found for Type-II.

\FloatBarrier
\subsubsection{Type-I Yukawa}
\label{sec:typeIyukawa}
If the couplings of the doublets to fermions are taken to be those of the Type-I 2HDM, i.e. taking all the charged fermions to couple only to $\Phi_2$, then the reduced couplings of the down- and up-type fermions to the CP-even scalars read
\begin{align}
    \kappa^{\mathcal{H}_a}_d = \kappa^{\mathcal{H}_a}_u= \frac{v}{v_2} x_{a2}\,,\ \ \  \ a\in\{1,2,3,4\}\,.
\label{eq:typeI_kappas-gen}
\end{align}
Specifying to $h_{125}$,
\begin{align}
    \kappa^{h_{125}}_d =   \kappa^{h_{125}}_u= \frac{v}{v_2} x_{22}\,
    \label{eq:typeI_kappas}
\end{align}
and considering, just as before, $\kappa^{h_{125}}_t = \kappa^{h_{125}}_b =1$, and using \cref{eq:typeI_kappas,eq:x23x24_relation,eq:h125eq3} along with $\sum_j x_{2j}^2   =1 $, we find 
\begin{equation}
v_1^2 + 4 u^2= \left(x_{21}^2 + 3 x_{24}^2\right) v^2\,,
\end{equation}
which indicates that the constraints are less stringent than those in the Type-II case, as $x_{21}$ remains free. On top of
this, $u$ can be raised even further if we allow a reasonable relaxation of $\delta=0$ and $\kappa^{h_{125}}_t=\kappa^{h_{125}}_b=1$, following the strategy described in \cref{app:A1}. In this case $|u|$ can exceed $\simeq 65$~GeV allowing for solutions with real-valued couplings. We will thus stick to Type-I in the rest of the paper.

\subsubsection{Determining the full sector}
\label{sec:full}
{Starting from the input strategy described in \cref{sec:strategy}, and taking into account the features discussed at the 
beginning of the present section and in \cref{sec:typeIyukawa,sec:typeIIyukawa,sec:typeXYyukawa} that favour Type-I 
over the three other types, we devise a well-defined procedure to determine the content of all four physical states 
$h_{95}, h_{125}, H_{320}$ and $H_{650}$ as well as their couplings to $W^+W^-$, $ZZ$ and $f\bar f$. 
We list hereafter its main steps, relegating more technical details to \cref{app:A1,app:A2,app:B}\footnote{We 
rely on the Mathematica package \cite{Mathematica} for the symbolic calculations and numerical scans as well as for the production of the figures.}.
\begin{enumerate}
    \item 
    We choose 6 inputs as discussed in \cref{sec:strategy}: (a)~$\kappa^{h_{125}}_W, \kappa^{h_{125}}_Z, \kappa^{h_{125}}_b$ 
    (b)~$\kappa^{h_{95}}_W, \kappa^{h_{95}}_t$, and    (c)~$\kappa^{H_{650}}_W$.
    A consistent choice of input numbers is described in the next subsection.
\item The input in (a) allows us to determine uniquely $v_1,v_2$ and the $x_{2i}$'s and to choose $u$ arbitrarily, albeit in a given domain of consistency; see \cref{app:A1} for details. 
\item The knowledge of the 4-vector with components $x_{2i}$ constrains the 4-vector $x_{1i}$ to live on a unit sphere in the hyperplane orthogonal to $x_{2i}$. Combined with the input in (b) and the first lines of \cref{eq:hww,eq:hzz}  and Eq.~\ref{eq:typeI_kappas}, the $x_{1i}$'s are fixed up to a possible
sign ambiguity; see \cref{app:A2} for details.
\item  The knowledge of the 4-vectors $x_{1i}$ and $x_{2i}$ constrains the 4-vector $x_{4i}$
to lie on a unit circle in the plane orthogonal to $x_{1i}$ and $x_{2i}$. Combined with the input in 
(c) and the fourth line of \cref{eq:hww}, this fixes the $x_{4i}$'s up to a discrete multiplicity; see \cref{app:A2} for details.
\item  From $x_{1i}$, $x_{2i}$ and $x_{4i}$ the 4-vector $x_{3i}$ is uniquely fixed up to a global sign, and all the remaining reduced couplings $\kappa^{h_{95}}_Z, \kappa^{H_{650}}_Z, \kappa^{H_{650}}_t, \kappa^{H_{320}}_W, \kappa^{H_{320}}_Z$ and  $\kappa^{H_{320}}_t,$ are predicted. 
\item Consistency with the various experimental indications can now be checked, in particular that of  $\kappa^{H_{650}}_W,\kappa^{H_{650}}_Z, \kappa^{H_{650}}_t$ with the correlation given by \cref{eq:kappa650corr}.
\end{enumerate}
}

\subsection{Results \label{sec:results}}
In the two following subsections, rather than presenting the results with the theoretical and experimental 
constraints all compiled in one step, we find it more instructive to somewhat break them up in order to 
understand their relative importance and effects. This is justified in particular since the present data from 
the LHC will change with future data taking.
\subsubsection{The general scan \label{sec:scan}}
\begin{table}[h]
\resizebox{.8\textwidth}{!}{
{
\begin{tabular}{|ll|}
  \hline
\hline
 $h_{125}:$ central values, cf. \cref{eq:CMS-SMkappas}~~~~~~~~~~~~~~~~~~~~~~~~& $\kappa_t^{h_{125}}=\kappa_b^{h_{125}}=0.99,\kappa_Z^{h_{125}}=1.04,\kappa_W^{h_{125}}=1.02$ \\
\hline\hline
 $h_{95}:$~~~~~~~~~~~~~~~~~~~~~~~~~~~~~~~~~~~~~~~~~~~~~~~~~~~~~~~~~~~~~~~~~~~~~~~~~~~ & scan on $\kappa_t^{h_{95}}=\kappa_\tau^{h_{95}}$ and $\kappa_W^{h_{95}}$  consistent with  \cref{eq:h95gamgam,eq:h95tautau}  \\
 \hline\hline
  \multirow{2}{20em}{~$H_{650}:$ cf. \cref{eq:WWH650min}} & $\kappa_W^{H_{650}}=0.89, 0.91$ with $c=0.85$ or $0.78$  \\
  & $\kappa_W^{H_{650}}=0.97, 1.0$ with $c=1.$   \\
 \hline\hline
 $H_{320}:$~~~~~~~~~~~~~~~~~~~~~~~~~~~~~~~~~~~~~~~~~~~~~~~~~~~~~~~~~~~~~~~~~~~~~~~~ &  $|\kappa_W^{H_{320}}|, |\kappa_Z^{H_{320}}| \lesssim 0.45$\\
 \hline
\end{tabular}}}
\caption{Inputs and constraints, assuming Type-I Yukawa.} \label{tab:inputs}
\end{table}

We show in this subsection the result of scans, taking as input the 6 reduced couplings
$\kappa^{h_{125}}_W, \kappa^{h_{125}}_Z, \kappa^{h_{125}}_b$, $\kappa^{h_{95}}_W, \kappa^{h_{95}}_t$, 
and $\kappa^{H_{650}}_W$ as indicated in Table \ref{tab:inputs}. 
\begin{itemize}
\item For $h_{125}$ we take the experimental 
central values for the couplings to $b\bar b$, $W^+W^-$ and $ZZ$ values as given in 
\cref{eq:CMS-SMkappas}. 
\item For $h_{95}$, we scan on $\kappa_t^{h_{95}}$ and $\kappa_W^{h_{95}}$ values that are consistent with 
the experimental indications, \cref{eq:h95gamgam,eq:h95tautau}, allowing 
one or two standard deviations from their central values; recall that here 
$\kappa_\tau^{h_{95}}=\kappa_b^{h_{95}}=\kappa_t^{h_{95}}$ since we are in the Type-I-2HDM-Yukawa scenario.  
Note that only the gauge boson and fermion virtual contributions to the $\gamma\gamma$ channel are included. 
We refer the reader to \cref{app:B} for details; let us just mention here that
charged and doubly-charged states can also contribute, however in a model-dependent way controlled by 
the couplings in the potential, a sector which we do not study in detail in the present paper. 
 Consideration of the charged scalars, however,
will not change the typical pattern presented in the figures and tables below.

Note that, given its controversial status \cite{Janot:2024ryq}, the LEP indication \cref{eq:h95LEP} for $h_{95}$ is {\sl not} systematically taken into account, but only shown as a further possible constraint referred to as LEPh95.

\item
For $\kappa_W^{H_{650}}$, we choose four representative values within the allowed narrow range given by \cref{eq:WWH650min} for three different values of $c$. 

\item
Since the indication for $H_{320}$ involves a substantial $H_{320} h_{125} h_{125}$ coupling (see the last line of 
\cref{tab:significance}), the couplings of $H_{320}$ to $ZZ$ and $WW$ should remain sufficiently small to account 
for the non-observation of decays to gauge bosons \cite{ATLAS:2022wti} . While this is difficult to quantify precisely, the last line of \cref{tab:inputs} should not be seen as an input motivated by quantitative experimental indications, but rather  as a qualitative guide for the scan 
presented in this section. We will however be brought to relax these bounds on $|\kappa_W^{H_{320}}|, |\kappa_Z^{H_{320}}|$ in the subsequent section.
\end{itemize}

\begin{figure}[htbp]
    \begin{center}
    \includegraphics[height=5cm]{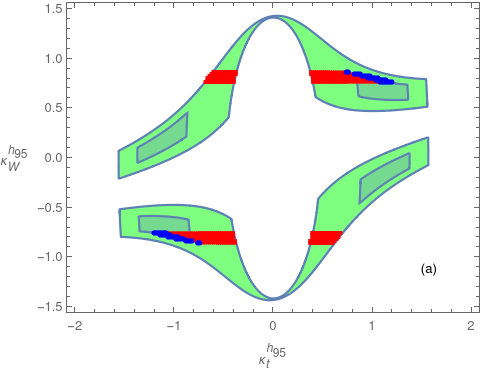}
    \includegraphics[height=5cm]{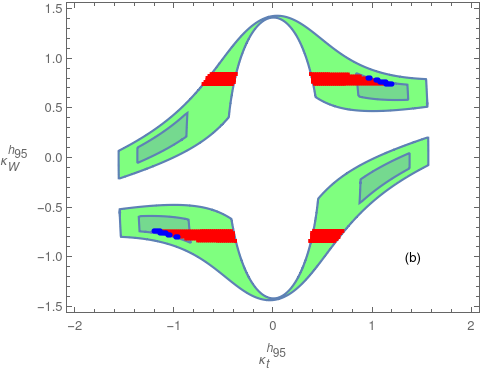}
    
\ \ \ \includegraphics[height=5cm]{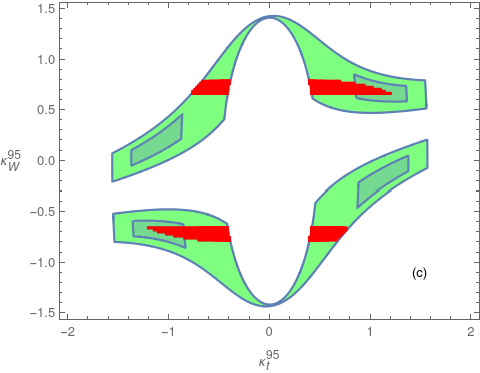}
 \includegraphics[height=5cm]{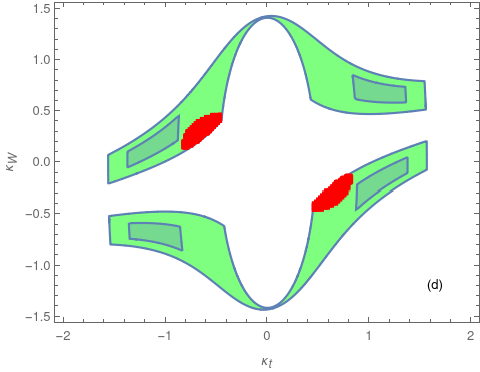}
    \end{center}
\caption{\small{{ $\kappa_t^{h_{95}}$ versus 
$\kappa_W^{h_{95}}$. The light (dark)  green regions correspond to the 2(1)$\sigma$ constraints \cref{eq:h95gamgam,eq:h95tautau}.
Fixing $\kappa_t^{h_{125}}=\kappa_b^{h_{125}}=.99,\kappa_Z^{h_{125}}=1.04,\kappa_W^{h_{125}}=1.02$, the red regions
correspond to requiring no complex-valued couplings or mixings, and varying the $H_{320}$ reduced couplings to W and Z  in the range $-.45\leq\kappa_W^{H_{320}}, \kappa_Z^{H_{320}} \leq +.45$; the blue regions correspond to overlaying the constraint given by \cref{eq:h95LEP} taken at the 2$\sigma$ level: (a) $\kappa_W^{H_{650}} =.89$, $u\simeq 78~{\rm GeV}, 
v_1 \simeq 16~{\rm GeV}, v_2 \simeq 76~{\rm GeV}$;
(b) $\kappa_W^{H_{650}} =.91$, $u\simeq 78~{\rm GeV}, 
v_1 \simeq 16~{\rm GeV}, v_2 \simeq 76~{\rm GeV}$;
(c) $\kappa_W^{H_{650}} =.97$, $u\simeq 78~{\rm GeV}, 
v_1 \simeq 16~{\rm GeV}, v_2 \simeq 76~{\rm GeV}$;
(d) $\kappa_W^{H_{650}} =1.$,   
$u\simeq 69~{\rm GeV}, v_1 \simeq 14~{\rm GeV}, v_2 \simeq 104~{\rm GeV}$.
}}}
\label{fig:ktkW-NEW}
\end{figure}


\Cref{fig:ktkW-NEW} shows the result of the allowed regions for $\kappa_t^{h_{95}} \left(=\kappa_b^{h_{95}}=\kappa_\tau^{h_{95}}\right)$ and $\kappa_W^{h_{95}}$, consistent with \cref{eq:h95gamgam,eq:h95tautau}, at the $1\sigma$ level (dark green) and $2\sigma$ level (light green). Obviously the green regions do not depend on the other inputs and are identical in the four panels of the figure. On top of that, the fixed 
$\kappa^{h_{125}}$s, and the real-valued nature of $\cal X$ and couplings, constrain the allowed ranges of the VEVs
$v_1,v_2$ and $u$ for each of the four chosen values of $\kappa_W^{H_{650}}$. Among the latter, the first three require $u$ at least as large as $\sim 78$~GeV, with $v_1 \simeq 16~{\rm GeV}, v_2 \simeq 76~{\rm GeV}$, corresponding to \cref{fig:ktkW-NEW}(a),(b),(c), while a somewhat smaller value $u\simeq 69~{\rm GeV}$ is obtained with $v_1 \simeq 14~{\rm GeV}, v_2 \simeq 104~{\rm GeV}$, for $\kappa_W^{H_{650}} =1$, as in \cref{fig:ktkW-NEW}(d). Even then, large fractions of the green regions are found not to be compatible with real-valued couplings of the remaining sectors. The red regions are the allowed ones where reality of $\cal X$ as well as the requirements of the last line of \cref{tab:inputs}  are fulfilled. Note also that only the smaller values of $\kappa_W^{H_{650}}$ allow for regions consistent with LEPh95, the blue regions in \cref{fig:ktkW-NEW}(a) and (b), while slightly larger $\kappa_W^{H_{650}}$ would not be consistent with the LEP indications. Thus, ignoring the latter conservatively, the red regions are the ones allowed by all theoretical constraints and experimental data.

\begin{figure}[htbp]
    \begin{center}
    \includegraphics[width=0.6\textwidth,keepaspectratio]{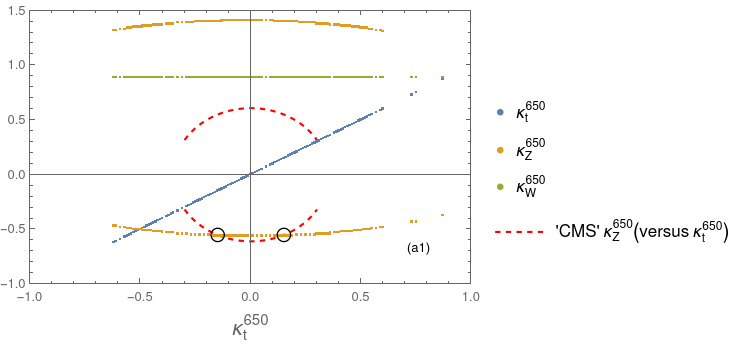}\includegraphics[width=0.4\textwidth,keepaspectratio]{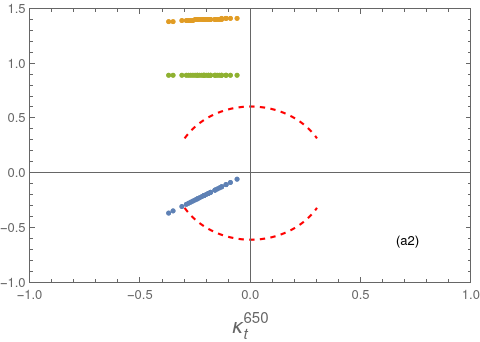}

    \includegraphics[width=0.6\textwidth,keepaspectratio]{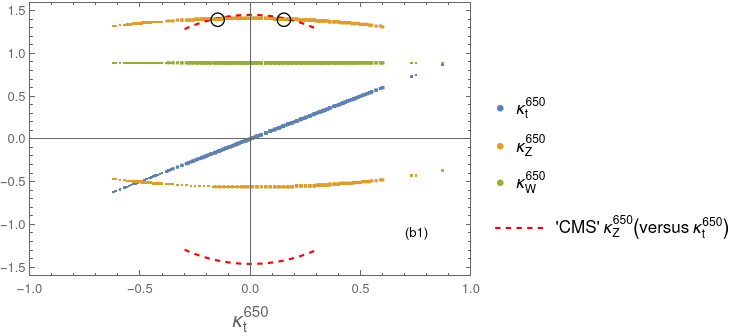}\includegraphics[width=0.4\textwidth,keepaspectratio]{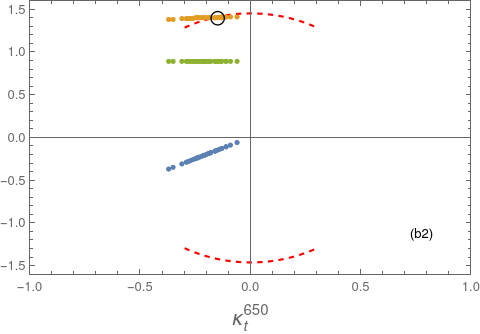}

    \includegraphics[width=0.6\textwidth,keepaspectratio]{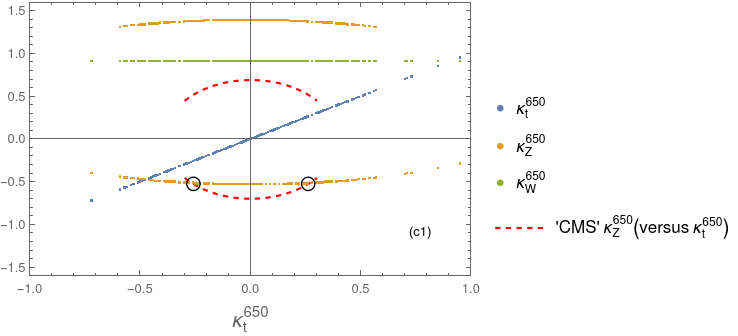}\includegraphics[width=0.4\textwidth,keepaspectratio]{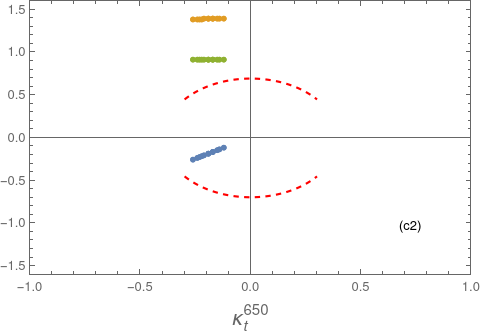}

    \includegraphics[width=0.6\textwidth,keepaspectratio]{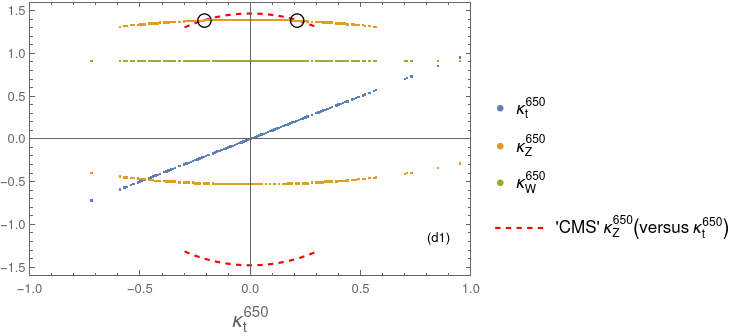}\includegraphics[width=0.4\textwidth,keepaspectratio]{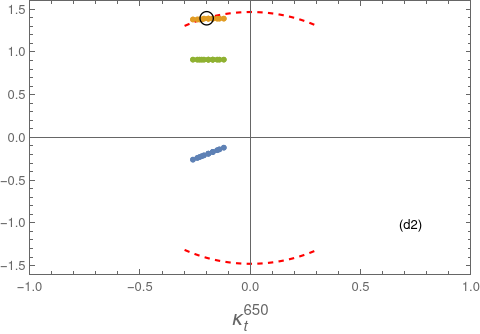}
        \end{center}
    \caption{\small{{Parameters and constraints taken from \cref{fig:ktkW-NEW},
    showing all $\kappa^{H_{650}}$ against $\kappa_t^{H_{650}}$. Figures~(a1), (a2), (b1), (b2) correspond to \cref{fig:ktkW-NEW}~(a): $\kappa_W^{H_{650}} =.89$, $u\simeq 78~{\rm GeV},
v_1 \simeq 16~{\rm GeV}, v_2 \simeq 76~{\rm GeV}$, with (a1), (a2): $c=.82$, (b1), (b2): $c=.75$.
Figures~(c1), (c2), (d1) ,(d2) correspond to \cref{fig:ktkW-NEW}~(b): $\kappa_W^{H_{650}} =.91$, $u\simeq 78~{\rm GeV}, v_1 \simeq 16~{\rm GeV}, v_2 \simeq 76~{\rm GeV}$, with
(c1), (c2): $c=.85$, (d1), (d2): $c=.78$. In figures~(a1), (b1), (c1), (d1)  all constraints are imposed except for LEPh95, \cref{eq:h95LEP}. Figures~(a2), (b2), (c2), (d2) correspond to overlaying the constraint given by LEPh95, \cref{eq:h95LEP},  taken at the 2$\sigma$ level. The black circles highlight 
the location of points that are also consistent with the CMS indication for $H_{650}$, \cite{CMS:2022bcb}. The corresponding full solutions are given in \cref{tab:kW650-.91} for the cases of figures (c1) and (d2).}}}
    \label{fig:ktkW650}
\end{figure}

\begin{figure}[htbp]
    \begin{center}
    \includegraphics[width=0.6\textwidth,keepaspectratio]{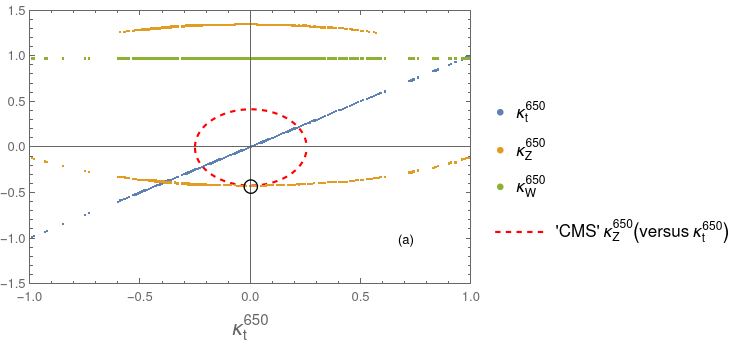}\includegraphics[width=0.4\textwidth,keepaspectratio]{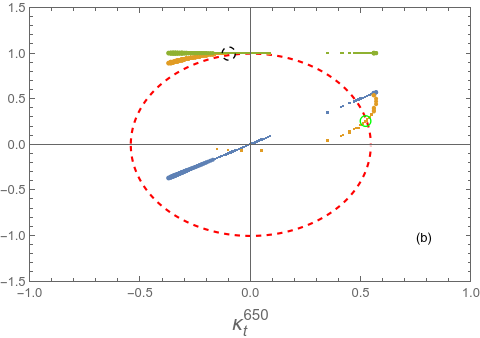}
    \end{center}
    \caption{\small{{Similar to
    \cref{fig:ktkW650}: parameters and constraints taken from \cref{fig:ktkW-NEW}. Figure (a) corresponds to \cref{fig:ktkW-NEW}~(c): $\kappa_W^{H_{650}} =.97$, $u\simeq 78~{\rm GeV},
v_1 \simeq 16~{\rm GeV}, v_2 \simeq 76~{\rm GeV}$.
Figure (b) corresponds to \cref{fig:ktkW-NEW}~(d): $\kappa_W^{H_{650}} =1.$, $u\simeq 69~{\rm GeV}, v_1 \simeq 14~{\rm GeV}, v_2 \simeq 104~{\rm GeV}$. For both cases $c=1$. The black circles highlight the location of points that are also consistent with the CMS indication for $H_{650}$, \cite{CMS:2022bcb}. None of these solutions satisfies LEPh95.}}}
\label{fig:ktkW650bis}
\end{figure}

Moving around in the allowed red regions, the construction of the ${\cal X}$ entries as described in \cref{sec:full} leads to the discrete branches for the reduced couplings of $H_{650}$ shown in \cref{fig:ktkW650,fig:ktkW650bis}. In particular, for a given fixed value of $\kappa_W^{H_{650}}$, there exists a 
one-branch solution for $\kappa_t^{H_{650}}$ but a two-branch solution for $\kappa_Z^{H_{650}}$. One can now bring in the correlation among 
$\kappa_Z^{H_{650}}, \kappa_W^{H_{650}}$ and $\kappa_t^{H_{650}}$, induced by the CMS indications,  \cref{eq:kappa650corr}. The ensuing $\kappa_Z^{H_{650}}$ as a function of $\kappa_t^{H_{650}}$  for each fixed value of $\kappa_W^{H_{650}}$ and illustrative choices of the normalisation factor $c$, is depicted by the red dashed lines in the figures. The final solutions are obtained as the overlap between this correlation and the lines of $\kappa_Z^{H_{650}}$. Examples are indicated by the black circles in \cref{fig:ktkW650,fig:ktkW650bis}.


\begin{table}[h]
{
{
\begin{tabular}{cccccccc}
\multicolumn{8}{c}{(A)} \\
  \hline
$\fbox{c=.78}$ & $\phi_1^0$ & $\phi_2^0$ & $\chi^0$ & $\xi^0$ & $\kappa_t$ & $\kappa_Z$ & $\kappa_W$ \\ \hline \hline
 $h_{95}$ & $0.35$ & $-0.45$ & $0.36$ & $-0.74$ & $-1.04$ & $0.47$& $-0.78$ \\
 $h_{125}$ & $0.73$ & $0.43$ & $0.44$ & $0.29$ & $0.99$ & $1.04$ & $1.02$ \\
 $H_{320}$ & $0.26$& $-0.77$ & $-0.06$ & $0.57$&  $-1.78$ & $-0.42$& $0.36$ \\
 $H_{650}$ & $-0.52$ & $-0.09$ & $0.82$ & $0.21$ & $-0.21$ & $1.39$ & $0.91$ \\
 \hline
 \multicolumn{8}{c}{\fbox{$u\simeq 78~{\rm GeV}, v_1 \simeq 16~{\rm GeV}, v_2 \simeq 76~{\rm GeV}$}}
 \end{tabular} }}
\resizebox{.86\textwidth}{!}
{
\begin{tabular}{cccccccc}
\multicolumn{8}{c}{(B)} \\
    \hline 
  $\fbox{c=.85}$ & $\phi_1^0$ & $\phi_2^0$ & $\chi^0$ & $\xi^0$ & $\kappa_t$ & $\kappa_Z$ & $\kappa_W$ \\ \hline \hline
 $h_{95}$ & $-0.62$& $0.3$& $0.72$& $0.03$& $0.68$& $1.37$& $0.76$ \\
 $h_{125}$ & $0.73$ & $0.43$ & $0.44$ & $0.29$ & $0.99$ & $1.04$ & $1.02$ \\
 $H_{320}$ & $0.12$& $-0.84$& $0.44$& $0.29$& $-1.94$& $0.43$& $0.4$ \\
 $H_{650}$ & $-0.25$& $0.12$& $-0.3$& $0.91$& $0.26$& $-0.52$& $0.91$ \\
 \hline
\end{tabular} 
\quad
{
\begin{tabular}{cccccccc}
\multicolumn{8}{c}{(C)} \\
  \hline
$\fbox{c=.85}$ & $\phi_1^0$ & $\phi_2^0$ & $\chi^0$ & $\xi^0$ & $\kappa_t$ & $\kappa_Z$ & $\kappa_W$ \\ \hline \hline
 $h_{95}$ & $0.41$ & $0.26$ & $-0.86$ & $-0.13$ & $0.6$ & $-1.39$ & $-0.78$ \\
 $h_{125}$ & $0.73$ & $0.43$ & $0.44$ & $0.29$ & $0.99$ & $1.04$ & $1.02$ \\
 $H_{320}$ & $-0.52$ & $0.86$& $0.01$& $0.02$& $1.96$& $0.34$& $0.36$ \\
 $H_{650}$ & $-0.16$ & $-0.11$& $-0.25$& $0.95$& $-0.26$& $-0.51$& $0.91$ \\
 \hline
 \end{tabular}}}
\caption{{Fractions of the scalar fields of the gauge basis as contained in the scalar fields of the mass basis, as well as the
reduced couplings to fermions and gauge bosons, for the case 
$\kappa_W^{H_{650}} =.91$, $u\simeq 78~{\rm GeV}, 
v_1 \simeq 16~{\rm GeV}, v_2 \simeq 76~{\rm GeV}$. Table (A) corresponds to the solution shown in \cref{fig:ktkW650}~(d2) that satisfies all constraints including LEPh95, \cref{eq:h95LEP}. Tables (B) and (C) correspond to the two solutions shown in \cref{fig:ktkW650}~(c1) that satisfy all constraints but LEPh95, \cref{eq:h95LEP}.
}}
\label{tab:kW650-.91}
\end{table}
\begin{table}[h]
\begin{center}
{
\begin{tabular}{cccccccc}
\multicolumn{8}{c}{(A)} \\
  \hline
$\fbox{c=1}$& $\phi_1^0$ & $\phi_2^0$ & $\chi^0$ & $\xi^0$ & $\kappa_t$ & $\kappa_Z$ & $\kappa_W$ \\ \hline \hline
 $h_{95}$ & $0.42$& $0.24$& $-0.87$& $-0.1$& $0.56$& $-1.41$& $-0.76$ \\
 $h_{125}$ & $0.73$& $0.43$& $0.44$& $0.29$& $0.99$& $1.04$& $1.02$ \\
 $H_{320}$ & $-0.48$& $0.87$& $0.03$& $-0.12$& $1.99$& $0.38$& $0.21$ \\
 $H_{650}$ &  $-0.24$& $0.$& $-0.23$& $0.94$& $0.$& $-0.43$& $0.97$ \\
 \hline
 \multicolumn{8}{c}{\fbox{$u\simeq 78~{\rm GeV}, v_1 \simeq 16~{\rm GeV}, v_2 \simeq 76~{\rm GeV}$}}
\end{tabular} \
\begin{tabular}{cccccccc}
\multicolumn{8}{c}{(B)} \\
  \hline
$\fbox{c=1}$& $\phi_1^0$ & $\phi_2^0$ & $\chi^0$ & $\xi^0$ & $\kappa_t$ & $\kappa_Z$ & $\kappa_W$ \\ \hline \hline
 $h_{95}$ & $0.24$& $-0.37$& $-0.37$& $0.82$& $-0.62$& $-0.8$& $0.42$ \\
 $h_{125}$ & $0.65$& $0.59$& $0.4$& $0.26$& $0.99$& $1.04$& $1.02$ \\
 $H_{320}$ & $-0.42$& $0.71$& $-0.52$& $0.21$& $1.19$& $-0.43$& $0.22$ \\
 $H_{650}$ & $-0.59$& $-0.01$& $0.66$& $0.47$& $-0.02$& $1.$& $1.$ \\
 \hline
 \multicolumn{8}{c}{\fbox{$u\simeq 69~{\rm GeV}, v_1 \simeq 14~{\rm GeV}, v_2 \simeq 104~{\rm GeV}$}}
\end{tabular}}
\end{center}
\caption{{Fractions of the scalar fields of the gauge basis as contained in the scalar fields of the mass basis, as well as the
reduced couplings to fermions and gauge bosons, satisfying all constraints except for LEPh95,  \cref{eq:h95LEP}: (A) corresponds to the solution $\kappa_W^{H_{650}} =.97$ shown in 
\cref{fig:ktkW650bis}(a); (B) corresponds to the solution $\kappa_W^{H_{650}} =1.$ shown in \cref{fig:ktkW650bis}(b).
}}
\label{tab:kW650-1}
\end{table}


In \cref{fig:ktkW650}, corresponding to \cref{fig:ktkW-NEW}(a) and (b), we compare the cases when LEPh95 is taken into account (the right panels) to the cases where this indication is ignored (the left panels). A common feature is that LEPh95 either eliminates the solutions altogether that are otherwise consistent with all the other indications, see \cref{fig:ktkW650} (a1) versus (a2) and (c1) versus (c2), or requires large values of $\kappa_Z^{H_{650}}$ that can cause tension with the hints for a 650~GeV object decaying to $ZZ$ \cite{ATLAS:2020tlo,ATLAS:2021kog}, \cref{fig:ktkW650} (b1), (b2), and (d1), (d2). Thus the LEPh95 indications appear to be only
marginally consistent, when all the other indications are considered together.

A given solution fixes the couplings and field content of the four scalar states. \Cref{tab:kW650-.91} shows the outcome for the full CP-even sector corresponding to the three solutions for the case $\kappa_W^{H_{650}} =0.91$.
\Cref{fig:ktkW650bis} and \cref{tab:kW650-1} show features similar to those discussed above, for the cases $\kappa_W^{H_{650}} =0.97$ and $\kappa_W^{H_{650}} =1.0$; these are, however, inconsistent with LEPh95. It is noteworthy that for all the solutions, the predictions for $\kappa_t^{H_{650}}$ come out naturally small to very small, in line with the CMS indications of suppressed ggF contributions \cite{CMS:2022bcb}. 

The solutions are, however, quite sensitive to the precise value of $\kappa^{H_{650}}_W$. Going from $0.97$ to 1 drastically changes the solutions for $\kappa^{H_{650}}_Z$, but $\kappa^{H_{650}}_t$ is found to be consistently small. On the other hand, $\kappa_t^{H_{320}}$ is predicted to be significantly large. This could be in tension with the experimental indications \cite{ATLAS:2022wti} of $H_{320}$ decaying mainly to a pair of SM-like Higgs boson. This would require large triple-scalar couplings to cope with the subdominance of the $t t^*$, $WW$ and $ZZ$ decay channels.

We have also checked the consistency of our benchmark points with {\tt HiggsTools} \cite{Bahl:2022igd}, and in particular, its sub-library
{\tt HiggsBounds-5} \cite{Bechtle:2020pkv}. 
To begin with, 
let us first drop the triple-scalar couplings and also forget the LEPh95 constraints. We find that three of the five
benchmark points in Tables \ref{tab:kW650-.91} and \ref{tab:kW650-1}, namely, III(B), III(C), and IV(A), remain 
allowed, while III(A) and IV(B) are ruled out by ATLAS \cite{ATLAS:2015pre, 
ATLAS:2018sbw} as well as CMS \cite{CMS-ZZ}, stemming from the large value of 
$\left|\kappa_Z^{H_{650}}\right|$, which results in an unacceptably large 4-lepton rate.

Now let us switch on the LEP constraints from negative searches for light scalars. This immediately rules out the
three surviving benchmark points III(B), III(C), and IV(A). 
This is due to the large $\left|\kappa_Z^{h_{95}}\right|$ values that exceed the experimental upper bound of $\simeq 0.48$ for an $h_{95}$ state with SM-like decays  \cite{LEPWorkingGroupforHiggsbosonsearches:2003ing,ADLO2}.
This is irrespective of whether one believes in the $h_{95}$ signal indications from LEP. 

These five benchmark points illustrate a tension between the LEP and LHC constraints due to a see-saw effect between $\kappa_Z^{H_{650}}$ and $\kappa_Z^{h_{95}}$ resulting from the orthogonality of the rotation matrix ${\cal X} in $\cref{eq:rotation}. However, one may 
note that such conclusions need not be true if the triple-scalar couplings, involving one $H_{650}$ state, are large enough to lead to sizeable decays to two lighter scalars. This would result in a smaller value of $\left|\kappa_Z^{H_{650}}\right|$, as discussed after \cref{eq:kappa650corr}, and thus in a smaller BR($H_{650}\to ZZ$). 
A detailed discussion of this issue follows now.

\subsubsection{Zooming in on viable solutions \label{sec:solutions}}
We just saw that, without the triple-scalar couplings, it is hard to find solutions consistent with both LEP and LHC. 
This is true not only for the benchmark points shown, but over the entire parameter space, the reason being that 
both $h_{95}ZZ$ and $H_{650}ZZ$ couplings are very tightly constrained. In this subsection, we will see what 
happens when a sizeable triple-scalar coupling is introduced. This is a two-step process. In the first step, we 
introduce triple-scalar couplings leading to di-scalar decays of $H_{650}$. 

The triple-scalar couplings can appear through both trilinear and quartic terms in the potential. 
Let us first start from the dimension-4 operator 
$\lambda\left(\Xi^\dag \Xi +X^\dag X \right) \left( \Phi_1^\dag\Phi_1 + \Phi_2^\dag \Phi_2\right)$, 
where $\lambda$ is a dimensionless coupling. Let us consider, more specifically, the decay 
$H_{320} \to h_{125} h_{125}$, originating, {\em e.g.}, from the term  
$\lambda X^\dag X \Phi_2^\dag\Phi_2$. Taking only the CP-even neutral component, shifting the vacuum and
rotating to the mass basis, the contribution to the trilinear coupling $H_{320} h_{125} h_{125}$ comes out
to be 
\begin{align}
H_{320} \to h_{125} h_{125}~:\ \ \ \ 
2\sqrt{2}\lambda  \left[ v_2\left(x_{23}^2 x_{32} + 2 x_{22} x_{23} x_{33}\right)+ u \left( x_{22}^2 x_{33} + 2 x_{22} x_{23} x_{32}\right)\right]\,.
\end{align} 
It is easy to see that all such terms must be of the form $\lambda$ times some VEV. As all the VEVs that we consider 
are less than 100 GeV, and the quartic couplings are expected to be perturbative, we may surmise that the coefficients
for the triple-scalar operators coming from the quartic terms in the scalar potential are typically less than 100 GeV.
 
On the other hand, one may have gauge-invariant trilinear terms of the form
\begin{align}
\frac{1}{\sqrt{2}} \left[M_\Xi  \Phi_i^\dag \bm{\tau}\Phi_j \cdot \Xi + 
M_X  \Phi_i^\top \sigma_2 \bm{\tau}\Phi_j \cdot \tilde{X} + 
M_{X \Xi} X^\dag \bm{t} X \cdot \Xi\right]\,,\ \ \ i,j\in \{1,2\}
\label{eq:trilin}
\end{align}
where the couplings $M_\Xi, M_X$ (taken here, for simplicity, the same for all three scalar doublet combinations) and $M_{X \Xi}$ have mass dimension one, and $\sigma_2$ is the second Pauli matrix\footnote{We refer the reader to \cite{KMP22} for the definitions of the $\bm{\tau}$ and $\bm{t}$ SU(2)-representation matrices in the spherical basis relevant for the vector representation of the triplets. The $\frac{1}{\sqrt{2}}$ normalisation factor allows direct matching with the couplings of the same operators in the matrix representation of the triplets \cite{Moultaka:2020dmb}.}.
If $M_X$, $M_\Xi$, and $M_{X\Xi}$ are at least of the order of 100 GeV or more, we may neglect, as a first approximation, 
the contributions coming from the quartic terms. For example, the $M_X$ term in Eq.\ (\ref{eq:trilin}) gives rise to  a factor of 
\begin{align}
    \frac{M_X}{\sqrt{2}} \left( x_{22}^2 x_{33} + 2 x_{22} x_{23} x_{32}\right)
\end{align}
for the same decay, namely, $H_{320}\to h_{125} h_{125}$, and this does not involve any VEVs. 
We will thus illustrate examples of consistent benchmark points in terms of the free couplings $M_\Xi, M_X$ and $M_{X \Xi}$.

\begin{figure}[htbp]
    \begin{center}
    \includegraphics[width=0.6\textwidth,keepaspectratio]{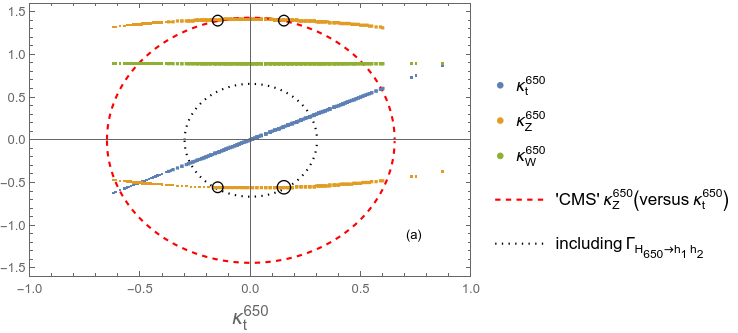}\includegraphics[width=0.4\textwidth,keepaspectratio]{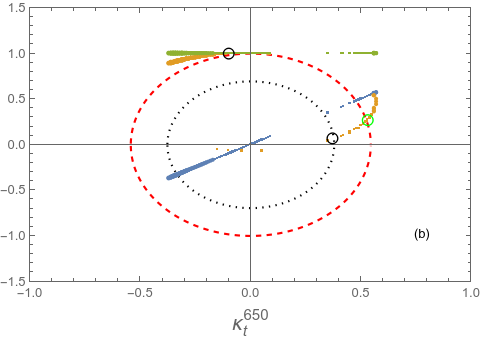}

    \includegraphics[width=0.65\textwidth,keepaspectratio]{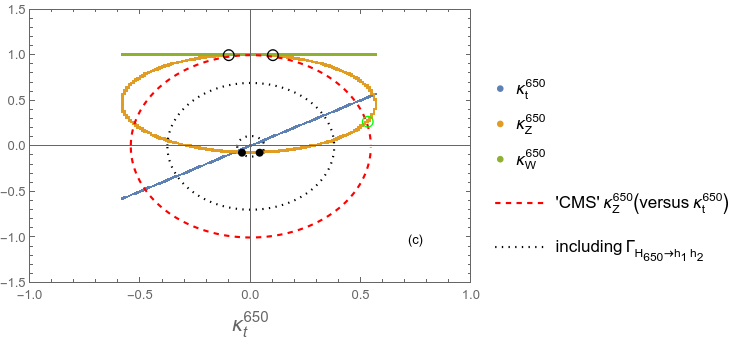}
    \end{center}
    \caption{\small{Figures (a) and (b) show the effect of $H_{650} \to h_{1} h_{2}$ decay: (a) taking 
$\Gamma_{H_{650} \to h_{1} h_{2}} \simeq 7.7$~GeV illustrates the modification to the solutions given in
    \cref{fig:ktkW650}(b1), where $\kappa_W^{H_{650}} =0.89$, $u\simeq 78~{\rm GeV},
v_1 \simeq 16~{\rm GeV}, v_2 \simeq 76~{\rm GeV}$ and $c=0.75$ ;
(b) taking $\Gamma_{H_{650} \to h_{1} h_{2}} \simeq 3.5$~GeV, illustrates the modification to the solution given in \cref{fig:ktkW650bis}(b), where
$\kappa_W^{H_{650}} =1.0$, $u\simeq 69~{\rm GeV}, v_1 \simeq 14~{\rm GeV}, v_2 \simeq 104~{\rm GeV}$ and $c=1$.
    Figure (c) illustrates the combined effect of relaxing the constraint $|\kappa_W^{H_{320}}|, |\kappa_Z^{H_{320}}| \lesssim 0.45$, cf. \cref{tab:inputs}, with the remaining input as in (b), and $H_{650} \to h_{1} h_{2}$ decays: $\Gamma_{H_{650} \to h_{1} h_{2}} \simeq 5$~GeV,
    (resp.\  $\simeq 3.5$~GeV),
    corresponds to the smaller (resp.\ larger) black-dotted ellipse. The two black blobs indicate the solutions given in \cref{tab:finalpoint}. See also the main text for further discussions.}
    }
    \label{fig:newpoints}
\end{figure}

Before considering explicitly the possible sources of $H_{650}$ decays to lighter scalars in the  2HDeGM, we stress first that the inclusion of such decay channels, although necessary,  would not be sufficient to delineate viable portions of the parameter space. We show, in \cref{fig:newpoints}(a) and (b), the effect of the two-body decays 
$H_{650} \to h_{1} h_{2}$,
where $h_{1,2}$ can be any of the three lighter scalar states. The red-dashed ellipses now shrink to the black-dotted ones
as a result of the substitution  in \cref{eq:kappa650corr} according to \cref{eq:shift}. The new solutions correspond to the intersections of the black-dotted ellipses with the orange lines, yielding indeed smaller values for $\left|\kappa_Z^{H_{650}}\right|$ than the ones initially obtained in \cref{fig:ktkW650}(b1) and \cref{fig:ktkW650bis}(b), and 
with ${\rm BR}^*_{H_{650} \to h_{1} h_{2}} \simeq 6\%$ and $3\%$ respectively, that remains sufficiently small to be consistent with the non-observation of these channels. However, this does not ameliorate the see-saw effect noted in the previous subsection. For instance the solution with $\kappa_Z^{H_{650}} \simeq -0.55$ in \cref{fig:newpoints}(a), and the solution with $\kappa_Z^{H_{650}} \simeq 0.07$ in \cref{fig:newpoints}(b), imply $\left|\kappa_Z^{h_{95}}\right| \simeq 1.4$ and $1.2$ respectively, values that are way above the LEP upper bounds.

We now go one step further. It turns out that to reduce the see-saw effects, one needs to relax the bounds 
shown in the last line of \cref{tab:inputs}. As stressed previously, these bounds on $\left|\kappa_Z^{H_{320}}\right|$ and  $\left|\kappa_W^{H_{320}}\right|$ are merely qualitative working assumptions. 
By relaxing them, one allows the see-saw effect to be shared among the $\kappa$s of the three states $H_{650}, H_{320}$ and $h_{95}$, and to populate regions where $\left|\kappa_Z^{h_{95}}\right|$ and $\left|\kappa_Z^{H_{650}}\right|$ take small values simultaneously, thus satisfying the experimental limits. This is illustrated in \cref{fig:newpoints}(c) which corresponds to the same input as for \cref{fig:newpoints}(b) but with $\left|\kappa_Z^{H_{320}}\right|$ and  $\left|\kappa_W^{H_{320}}\right|$ allowed to vary from 0 to 1.5. The orange line is now a complete ellipse and viable solutions can be obtained for moderately small widths for $H_{650} \to h_{1} h_{2}$. As an example, assuming a width of $\simeq 5$~GeV, that corresponds to ${\rm BR}^*_{H_{650} \to h_{1} h_{2}} \lesssim 6\%$, and restricting to $\left|\kappa_t^{H_{650}}\right|, \left|\kappa_t^{H_{320}}\right| \lesssim 0.1$, one finds the two benchmark points,
albeit very close in the parameter space (because the space is so tightly constrained), given in \cref{tab:finalpoint} for which the $h_{95}$ and $H_{650}$ states satisfy all the corresponding  constraints discussed so far in the paper (on top of those for the SM-like $h_{125}$ state). 

\begin{table}[h]
\begin{center}
{
\begin{tabular}{||cccccccc||}
\multicolumn{8}{c}{(A)} \\
  \hline
$\fbox{c=1}$ & $\phi_1^0$ & $\phi_2^0$ & $\chi^0$ & $\xi^0$ & $\kappa_t$ & $\kappa_Z$ & $\kappa_W$ \\ \hline \hline
 $h_{95}$ & $-0.48$ & 0.8 & $-0.27$ & $-0.22$ & 1.34 & 0.01 & $-0.02$ \\
 $h_{125}$ & 0.65 & 0.59 & 0.4 & 0.26 & 0.99 & 1.04 & 1.02 \\
 $H_{320}$ & $-0.46$ & $-0.02$ & 0.88 & $-0.15$ & $-0.03$ & 1.35 & 0.48 \\
 $H_{650}$ & $-0.37$ & 0.02 & $-0.03$ & 0.93 & 0.03 & $-0.07$ & 1.0 \\
 \hline
\end{tabular}
\begin{tabular}{||cccccccc||}
\multicolumn{8}{c}{(B)} \\
  \hline
$\fbox{c=1}$ & $\phi_1^0$ & $\phi_2^0$ & $\chi^0$ & $\xi^0$ & $\kappa_t$ & $\kappa_Z$ & $\kappa_W$ \\ \hline \hline
 $h_{95}$ & $-0.47$ & 0.8 & $-0.31$ & $-0.17$ & 1.34 & $-0.06$ & 0.0 \\
 $h_{125}$ & 0.65 & 0.59 & 0.4 & 0.26 & 0.99 & 1.04 & 1.02 \\
 $H_{320}$ & $-0.48$ & 0.02 & 0.86 & $-0.16$ & 0.03 & 1.35 & 0.48 \\
 $H_{650}$ & $-0.35$ & $-0.02$ & $-0.02$ & 0.94 & $-0.03$ & $-0.07$ & 1.0 \\
 \hline
 \end{tabular}}\\
{
 \begin{tabular}{cccccccc}
 \multicolumn{8}{c}{\fbox{$u\simeq 69~{\rm GeV}, v_1 \simeq 14~{\rm GeV}, v_2 \simeq 104~{\rm GeV}$}}
\end{tabular}}
\caption{\small{New benchmark points assuming  ${\rm BR}^*_{H_{650} \to h_{1} h_{2}} \lesssim 6\%$, requiring $\left|\kappa_t^{H_{650}}\right|, \left|\kappa_t^{H_{320}}\right| \lesssim .1$ and relaxing 
$|\kappa_W^{H_{320}}|, |\kappa_Z^{H_{320}}| \lesssim 0.45$, 
for which the $h_{95}$ and $H_{650}$ constraints are satisfied. See main text for further discussions.}}
\label{tab:finalpoint}
\end{center}
\end{table}

The second step is not yet complete. It still remains to be seen whether $H_{320}$ complies with the corresponding experimental indications, in particular since now 
$\left|\kappa_Z^{H_{320}}\right|$ and  $\left|\kappa_W^{H_{320}}\right|$ are allowed to have large values.  Taking the numbers in  \cref{tab:finalpoint} as a benchmark and using {\tt HiggsBounds-5} \cite{Bechtle:2020pkv}, we find that the total decay width of $H_{320}$ should be increased by 250\% to satisfy the experimental limits on VBF-produced objects decaying to a pair of Z bosons, \cite{ATLAS:2016oum}, while decays to W's and fermions remain within the experimental limits. This means a decay branching fraction into final states, other than gauge boson and fermion pairs, of about 70\%, which is in accordance with the experimental indications of decays to pairs of $h_{125}$, see \cref{tab:significance}. One will however have to cope with exclusion limits applicable to
both $H_{650}$ and $H_{320}$, from searches for heavy scalars decaying to a SM-like Higgs and a lighter scalar, \cite{CMS:2021yci} and \cite{ATLAS:2025nda}.

The  trilinear couplings among the physical states, $H_{650}h_a h_b$ and $H_{320}h_a h_b$, where 
$a,b=95$ and/or $125$, are uniquely fixed in terms of the $x_{ij}$'s and the three mass parameters. 
Moreover, the production cross-sections for $H_{650}$ and $H_{320}$, as well as their decays to fermions and gauge bosons, are fixed by the knowledge of the corresponding reduced couplings $\kappa$ to fermions and gauge bosons.
We have computed the branching ratio BR$[\displaystyle H_{650} \to \sum_{a,b} h_a h_b]$ of the 
$H_{650}$ total decay to lighter scalars, requiring it to be $\lesssim 6 \%$ that corresponds to the solution given in \cref{fig:newpoints}(c). We also computed the production cross-sections times branching ratios $\sigma\left(pp \to H_{650} \to h_{95}(b\bar{b}) h_{125}(\tau^+ \tau^-)\right)$ and $\sigma\left(pp \to H_{320} \to h_{95}(b\bar{b}) h_{125}(\tau^+ \tau^-)\right)$, akin to the limits from Fig.\ 5 of Ref.\ \cite{CMS:2021yci}, and $\sigma\left(pp \to H_{650} \to h_{95}(b\bar{b}) h_{125}(\gamma \gamma)\right)$ and $\sigma\left(pp \to H_{320} \to h_{95}(b\bar{b}) h_{125}(\gamma \gamma)\right)$, akin to the limits from Fig.\ 3 of Ref.\ \cite{ATLAS:2025nda}. 
Specifying to the benchmark points of \cref{tab:finalpoint}
we find that all the experimental constraints discussed above can be satisfied, the one for $\sigma\left(pp \to H_{320} \to h_{95}(b\bar{b}) h_{125}(\gamma \gamma)\right)$ being the most stringent. Taking a conservative upper bound of $\simeq 0.8$~fb  for the latter \cite{ATLAS:2025nda}, we find,
\begin{align}
115~{\rm GeV} \lesssim M_X \lesssim 150~{\rm GeV}\,,\ \ \ 
\left|M_\Xi\right| \lesssim 150~{\rm GeV}\,,\ \ \  
400~{\rm GeV}\lesssim M_{X \Xi} \lesssim 500~{\rm GeV}\,,
\label{eq:Mbounds}
\end{align}
a range compatible with the physical masses of the four scalars. Relaxing the $\sigma\left(pp \to H_{320} \to h_{95}(b\bar{b}) h_{125}(\gamma \gamma)\right)$ bound to,
say, $1.5$~fb opens up several disjoint domains in the
$(M_X, M_\Xi, M_{X \Xi})$ space, the one compatible with the four physical masses being enlarged to
\begin{align} 
200~{\rm GeV} \lesssim M_X \lesssim 430~{\rm GeV}\,,\ \ \  
 150~{\rm GeV} \lesssim M_\Xi \lesssim 300~{\rm GeV}\,,\ \ \ 
200~{\rm GeV}\lesssim M_{X \Xi} \lesssim 560~{\rm GeV}\,.
\end{align}

\begin{table}[h]
\resizebox{.7\textwidth}{!}{
{
\begin{tabular}{cccccc}
  \hline
Benchmark & $\sigma^{VBF}_{pp \to H_{650} j j}$(fb) & $\Gamma_{H_{650} \to W^+ W^-}$(GeV) & $\Gamma_{H_{650} \to Z Z}$(GeV) & $\Gamma_{H_{650} \to t \bar{t}}$(GeV) & $\Gamma_{H_{650} \to \sum_{a,b} h_a h_b}$(GeV)\\ \hline \hline
  V(A) & $124$ & $81.1$ & $0.19$ & $0.02$ & $5$ \\
 \hline
 \end{tabular} }}
 \caption{The VBF production cross-section (see also footnote  \ref{footnote:6}) and main decay widths of $H_{650}$, for the benchmark point of \cref{tab:finalpoint} (A) corresponding to the solution given in \cref{fig:newpoints}(c). The illustrated value of the total decay width to lighter scalars, $\Gamma_{H_{650} \to \sum_{a,b} h_a h_b}$, corresponds to $M_X \simeq 148$~GeV, $M_\Xi \simeq -60$~GeV and $M_{X \Xi} \simeq 500$~GeV,  consistent with the conservative bounds given by \cref{eq:Mbounds}.}
 \label{tab:Xsection}
 \end{table}

Finally, we give in \cref{tab:Xsection} the production cross-section, which is dominated by the VBF process, and the main decay widths of $H_{650}$, resulting from the benchmark points of \cref{tab:finalpoint} that are consistent with all the constraints.

In summary, the discussion in this subsection has shown why the allowed space is so tightly constrained. Based on this discussion, one can now see pictorially from \cref{fig:ktkW650} and \cref{fig:ktkW650bis}~(a) why none of these configurations can give viable solutions even when including decays to scalars and/or relaxing  the last constraint of \cref{tab:inputs}. These configurations correspond to the three sets of values of $u,v_1,v_2$ and
  $\kappa_W^{H_{650}}$ corresponding to \cref{fig:ktkW-NEW}~(a), (b) and (c). 
  {\color{blue} The rationale for discussing these benchmark points is to show explicitly how the parameter space
  gets constrained by data.} 
  We are thus left with the set of values
  $u\simeq 69~{\rm GeV}, v_1 \simeq 14~{\rm GeV}, v_2 \simeq 104~{\rm GeV}$ and $\kappa_W^{H_{650}}=1$ corresponding to
  \cref{fig:ktkW-NEW}~(d) that lead to the benchmark points
  of \cref{tab:finalpoint}, thus determining allowed regions in the vicinity of these points. However, one should not conclude that the only consistent solutions are restricted to these regions. For instance, allowing now $\kappa_t^{H_{650}}$ and $\kappa_t^{H_{320}}$  to lie in the range $.1 \lesssim \left|\kappa_t^{H_{650}}\right|, \left|\kappa_t^{H_{320}}\right| \lesssim .2$, one finds  several other benchmark points with the latter $\left|\kappa_t\right|$'s ranging from $.14$ to $.19$ and acceptable regions around these points, each of them leading to acceptable ranges for $M_X, M_{\Xi}$ and $M_{X \Xi}$, similar to the ones discussed above. It remains true, though, that the compilation of all available constraints reduces drastically the possibility of fitting simultaneously the four scalar resonances.  In particular, very stringent constraints originate from requiring small $\kappa_Z^{H_{650}}$ and $\kappa_Z^{h_{95}}$, due respectively to the present limits on 4-lepton events  from heavy resonance decays, and to limits on Higgsstrahlung production of light scalars decaying to fermions.

\subsection{Indirect constraints}
For the indirect constraints, like those coming from flavour physics, or muon $(g-2)$, one needs
the fermionic couplings of the scalars. Only the SU(2) doublets couple with fermions (if we
neglect the lepton-number violating coupling of the triplets). Thus, one expects the flavour
constraints to be typically close to those coming from the different two-Higgs doublet models.

Muon $(g-2)$ shows a tension of more than $5\sigma$
with the SM expectation \cite{ParticleDataGroup:2024cfk}:
\begin{equation}
\Delta a_\mu = a_\mu^{\rm exp} - a_\mu^{\rm SM} = \left[ 24.9 \pm 2.2 (\mbox{expt.}) \pm
4.3 (\mbox{theo.}) \right] \times 10^{-10}\,.
\end{equation}
The dominant contribution from extra neutral scalars is the one-loop magnetic moment
amplitude, mediated by CP-even or CP-odd scalars. A similar contribution comes from the
charged Higgs mediated diagram too. The coupling is proportional to $\tan\beta$
for Type-II and Type-X models, and $\cot\beta$ for Type-I and Type-Y. Thus, we expect
the tension $\Delta a_\mu$ to go down for Type-II and Type-X models, and remain the same for 
Type-I and Type-Y
models, if we work in the large $\tan\beta$ limit. One must remember that in 2HDeGM, components of
$\Phi_1$ and $\Phi_2$ are distributed among all the physical scalars, some of them being quite heavy, 
so we do not expect as large a contribution as in 2HDMs. 

The most important constraint from quark flavour physics is on the mass of the charged Higgs, $m_{H^+}$,
that comes from $b\to s\gamma$. The charged Higgs contribution tends to increase the difference between 
the theoretical prediction from SM and the experimental number, and so gives 
a very tight constraint on the $H^+$ mass when such contributions are significant. 
For 2HDM Type-II and Type-Y, Ref.\ \cite{Misiak:2017bgg} finds an absolute limit of $m_{H^+} > 583$ GeV at
95\% confidence limit, which is valid for all $\tan\beta$. Bounds from Type-I and Type-X are much weaker; for 
$\tan\beta=1$, one finds $m_{H^+} > 445$ GeV at 95\% confidence limit, but the lower limit
drops sharply with increasing
$\tan\beta$. Thus, for these two types of 2HDM, there is essentially no constraint from flavour physics in the 
large $\tan\beta$ limit. A similar conclusion holds for $B_s\to\mu^+\mu^-$. 

Other indirect constraints come from Higgs signal strengths and oblique parameters. 
Authors of Ref.\ \cite{Chowdhury:2017aav}
performed a global fit of 2HDM by combining all theoretical and experimental bounds and presented the 
allowed parameter space. The latest Run II signal strength data from LHC have been imposed and the results 
show that the deviation from the alignment limit $\beta - \alpha = \pi/2$ cannot be larger than 0.26, 0.055, 0.069 and 
0.056 for 2HDM Types I, II, X and Y respectively at $2\sigma$ confidence level. 
The fit includes electroweak precision observables too.

One may wonder whether the constraints on the parameter space of the canonical GM model 
\cite{Georgi:1985nv}, like those on the triplet VEV $u$, are still valid. For example, Ref.\ \cite{Hartling:2014aga}
finds $u$ to be about 40 GeV or less, mostly from $b\to s\gamma$. From a global fit, the authors of Ref.\
\cite{Chiang:2018cgb} draw a similar conclusion. We have consistently worked with a larger value of $u$. One may
note that not only our model has more fields than canonical GM, but there is also no degeneracy among the member
of the custodial multiplets. The latter fact opens up new decay channels, like that of the doubly charged Higgs, and
so the constraints coming from the LHC search on $H^{++}\to W^+W^+$ need not be valid any longer. The former fact
may also allow for possible cancellation among different radiative corrections.

\section{Further ramifications}     \label{sec:moretheory}
\subsection{Charged scalars}     \label{sec:chargedscalars}
2HDeGM is a minimalistic model to explain all the indications, and it can be pruned down further if some of the 
indications go away. If they all persist, one would like to ask where the charged scalars and the CP-odd scalars are. 

We note that there are some mild hints. For example, the CMS Collaboration \cite{CMS:2021wlt}
shows a mild excess for $H^+\to W^+Z\to 3\ell+\nu$ 
at $m_{H^+}\sim 375$ GeV and another mild excess for $H^{++}\to W^+W^+\to 2\ell + e\nu$ at $m_{H^{++}} \sim
450$ GeV. They do not quote any numbers. Surprisingly, ATLAS also found a mild signal of $H^+$ at 375 GeV 
\cite{ATLAS:2022zuc}, where the production is through VBF, with a local (global) significance of $2.8\, (1.6)
\sigma$. They also found \cite{ATLAS:2023dbw} a $3.2 \, (2.5) \sigma$ local (global) significance signal for 
$H^{++}$ at 450 GeV, decaying to $W^+W^+$.

There are indeed very strong bounds on the mass of $H^{++}$ of the canonical GM model from the 
CMS Collaboration \cite{CMS:2021wlt}. From the non-observation of $H^{++}\to W^+W^+$, a lower bound of $2~{\rm TeV} \lesssim m_{H^{++}}$ has been inferred. Similarly, the non-observation of $H^+\to W^+Z$ gives a lower bound
$1.1~ {\rm TeV} \lesssim m_{H^+}$. They have been obtained for $\sin\theta_H=1$, while the VBF cross-section 
and hence the total number of events depend on $\sin^2\theta_H$. Even without taking that
into account, we have seen in the previous Section 
that such bounds are not watertight. For example, in 2HDeGM where 
the scalar custodial multiplets are not degenerate, $H^{++}$ can decay to $H^+W^+$, where $H^+$ is some
singly charged scalar of the model. It can also decay to two singly charged scalars. While the former branching ratio
can be estimated from the knowledge of the charged Higgs mass matrix ({\em i.e.}, the components of the 
custodial 5-plet singly charged scalar in different mass eigenstates), the latter needs a knowledge of the scalar 
potential and hence cannot even be estimated at this stage. What can definitely be said is that the CMS bounds 
should not hold any longer and the mass of $H^{++}$ can indeed be less than 1 TeV.

A similar conclusion holds for all the singly charged scalars. If they are above the $t\bar{b}$ threshold and have 
a significant SU(2) doublet component, they should be searched in the $H^+\to t\bar{b}$ channel; otherwise, 
one has to look for $H^+\to c\bar{b}, \tau^+\nu_\tau$. If the charged scalars are sufficiently apart, {\em e.g.}, 
one close to 375 GeV and the other, say, at about 130 GeV,
there can be a cascade like $H_1^+\to H_2^++Z$.

\subsection{CP-odd neutral scalars}   \label{sec:cpodd}
One possible CP-odd scalar is $A_{400}$. For the CP-odd scalars, there is a different mixing matrix between the 
mass and the gauge eigenstates, one of whose rows is determined from the fact that among the three CP-odd 
scalars in the theory, one has to be the neutral Goldstone boson. To determine the matrix, it is important to
get the rate for $A_{400}\to t\bar{t}$. One may note that the production must proceed through ggF, or associated
production with $t\bar{t}$. 

2HDeGM can accommodate two CP-odd neutral scalars, and hopefully the second slot will be taken up
by $A_{151}$ when it is confirmed by the experimentalists.
Since these are broad resonances, there is also the possibility that two or more resonances are closely spaced and 
cannot be resolved. There is also the possibility that the scalar potential is not CP-invariant, and hence the mass 
eigenstates are no longer CP-eigenstates. These issues may all be resolved in the next $e^+e^-$ or muon collider. 

\section{Summary and conclusion}    \label{sec:conclude}
Over the last few years, there have been strong hints of more resonances from the LHC experiments,
at around 95~GeV and 650~GeV, and possibly at 320~GeV and 400~GeV, 
even though none of them are officially confirmed as yet.
Assuming that these are all scalar resonances, we try to accommodate them, 
together with the SM-like 125~GeV resonance, in a minimalist model.
As far as we know, this is the first
attempt to include all such hints. To be conservative, we take into account only the
significances announced by the experimental collaborations.
We thus find the combined global significance for the 95~GeV resonance to be around $3\sigma$,
while that for the 650~GeV resonance has reached the 4$\sigma$ level. Another resonance at around 400 GeV,
which is in all probability a CP-odd neutral scalar, has also breached the $3\sigma$ mark, and the significance of 
the one at 320 GeV is just below $3\sigma$.

We first show that the coupling strengths of these new scalars to the weak gauge bosons predict
the existence of doubly charged scalars, if the underlying theory is renormalisable and
the unitarity bounds respected. This, in turn, leads us to discard the simplest scalar extensions
like those with $\rm SU(2)_L$ singlets and doublets, and help us to arrive at the minimal model, with
two $\rm SU(2)_L$ doublets, one real triplet, and one complex triplet. The triplet structure is similar to
the Georgi-Machacek model, with identical VEVs for the triplet fields to ensure $\rho=1$ at the tree level,
except that the custodial symmetry should not be respected in the scalar sector
and hence the custodial multiplets are no longer mass degenerate. We call this the 2-Higgs Doublet extended
Georgi-Machacek (2HDeGM) model. This model accommodates not only all the neutral scalars,
both CP-even and CP-odd, but also singly and doubly charged scalars for which there are some mild hints
from both ATLAS and CMS Collaborations.

While the model may seem to contain a lot of fields and free parameters, and the amount of data may seem 
scant, we note with pleasant surprise that even the existing
data, taken at face value, severely constrain the allowed parameter space, without even going into
a detailed discussion of the scalar potential and resulting constraints. 
This, in turn, indicates very narrow ranges for the couplings of the 650~GeV scalar $H_{650}$ to the
electroweak gauge bosons.
This also indicates a large di-scalar decay width for $H_{320}$, which is still allowed from the CMS and ATLAS analyses,
but more data and/or a more sensitive analysis may easily rule this model out.

We assume CP to be conserved in the scalar sector and all mass eigenstates to be CP eigenstates also.
Within the narrow ranges of the couplings, rotating the four CP-even states to their mass eigenstates 
entails theoretical consistency constraints that cut down the possibilities to very few solutions. 
Let us summarise the major characteristics of the solutions:
\begin{itemize}
    \item Over most of the parameter space, defined in terms of the masses, or the parameters of the scalar potential, 
    there exists irreconcilable tension among the various observables, coming mostly from the LHC. Two of the most 
    important constraints are the paucity of events in the 4-lepton channels coming from the decay of $H_{650}$, and the 
    constraints on 
    a light Higgs at 95 GeV produced through Higgsstrahlung and decaying into light fermions, that put 
    strong upper bounds on $\kappa_Z^{H_{650}}$ and $\kappa_Z^{h_{95}}$. While $\kappa_Z^{H_{650}}$ has to be
    small, the couplings of $H_{650}$ with $W^+W^-$ can be large (and have to be large to be consistent with the CMS
    indications). 
    \item The orthogonality of the ${\cal X}$ matrix makes the implementation of the above condition difficult. We find 
    that we must allow for a large di-scalar decay width for $H_{320}$, in all probability to $h_{95} h_{125}$. 
    \item There are many ways to refine the analysis, but almost all of them requires a thorough knowledge of the scalar 
    potential, which we do not have. Even then, it is interesting to note that we can put nontrivial constraints on the 
    allowed values of the trilinear scalar couplings. The decay $h_{95}\to\gamma\gamma$ can also be modified by singly- and 
    doubly-charged scalars in the loop, although such effects are expected to be small. 
    \item We also find that, under conservative assumptions about the signs 
of the SM-like Yukawa couplings, Type-I Yukawa structure is favoured over the three other conventional types.\end{itemize}

Given the large number of searches for new resonances by ATLAS and CMS, few excesses above 2$\sigma$ 
are bound to occur, eventually disappearing with more statistics. Nonetheless, in the absence of robust 
theoretical guide to where physics beyond the SM, if any, might show up, the interpretations of excesses 
remains a precious tool that might shed light on the right way to go. Our study illustrates possibilities and 
issues when trying to interpret simultaneously several excesses within a given theoretical model. It also 
sets the stage for further investigations depending on future data. 
For example, (i) $H_{650}\to W^+W^-$ and/or $ZZ$ , and also $t\bar{t}$ 
if backgrounds can be controlled, (ii) $H_{320}\to W^+W^-, ZZ$, and (iii) $h_{95} \to 
\tau^+\tau^-$ in associated production with the top, should be prioritized, while one should 
also look for a heavy scalar decaying to two light scalars, or a light scalar and a weak gauge boson.
We would also like to mention that this is an exploratory exercise that illustrates the difficulties
when one tries to fit several resonances simultaneously, even for next-to-minimal extensions of the
SM. A further extension of the model that we propose should be much less constrained, but that is
outside the scope of this paper.

\begin{acknowledgments}
 We would like to thank Fran\c{c}ois Richard for many helpful comments and suggestions, and for
    drawing our attention to the large significance of $H_{650}$, and Alain Le Yaouanc for fruitful discussions.
    We also thank Sunando K.\ Patra for bringing Ref.\ \cite{theiler} to our attention. 
A.K.\ acknowledges ANRF, Government of India, for support through the grant CRG/2023/000133. P.M.\ acknowledges ANRF, Government of India, for support through the
grant CRG/2021/007579.
G.M. acknowledges partial support from the European Union’s Horizon 2020 research and innovation programme under the Marie Skłodowska-Curie grant agreements No 860881-HIDDeN and No 101086085–ASYMMETRY.
\end{acknowledgments}

\appendix

\section{Determining $v_1,v_2,u$ and the $x_ {ij}$s}
\subsection{Information from $h_{125}$  \label{app:A1}}
The reduced couplings $\kappa_W^{h_{125}}$ and $\kappa_Z^{h_{125}}$    are defined
as follows:
\begin{equation}
 g_{h_{125}WW} = g^2 \frac{v}{\sqrt{2}} \kappa_W^{h_{125}}, \ \ g_{h_{125}ZZ} = g^2 \frac{v}{2\sqrt{2}\cos^2\theta_W} \kappa_Z^{h_{125}},
\end{equation}
where $g_{h_{125}WW}$ and $g_{h_{125}ZZ}$ denote the true couplings. The SM couplings are retrieved in the limit $\kappa_W^{h_{125}}=\kappa_Z^{h_{125}}=1$ and taking
$\cos^2\theta_W = M_W^2/M_Z^2$ at the tree-level.

Let us assume $\kappa_W^{h_{125}}$ and $\kappa_Z^{h_{125}}$
are extracted from experiment (ignoring experimental errors at this
stage) and try to determine $v_1, v_2$ and $u$
in general, that is without necessarily adopting
the {\em alignment limit} $h_{125}\approx h_{\rm SM}$. 
Of course this limit will
be a special case of the general solutions.
One reason for trying the general solution is to fully
control the value of $u$ which is essential for
the discussion of the constraints on the doubly-charged Higgs.
Furthermore, once the solutions are obtained, experimental errors can
be easily propagated.

\Cref{eq:hww,eq:hzz} give the theoretical expressions of the scalar to $W^+W^-$ and $ZZ$ couplings, with,
 for the latter, an extra symmetry factor of two to account for the identity of the two Z bosons.
 After canceling some common factors, the relevant expressions for the reduced couplings read,
\begin{align}
v \kappa^{{\cal H}_{a}}_{W} =& 
 v_1 x_{a 1} + v_2 x_{a 2} + 2 u (x_{a 3} + 
 \sqrt{2} x_{a 4}), \label{eq:WWS}\\
 v \kappa^{{\cal H}_{a}}_{Z} =& v_1 x_{a 1} + v_2 x_{a 2} + 4 u x_{a 3}, \  \label{eq:ZZS}
\end{align}
where $a=1,...,4$. 
Let us  consider in particular the reduced couplings involving $h_{125}$:
\begin{align}
  v \kappa_W^{h_{125}} =& 
 v_1 x_{2 1} + v_2 x_{2 2} + 2 u (x_{2 3} + 
 \sqrt{2} x_{2 4}), \label{eq:h125eq1}\\
v \kappa_Z^{h_{125}} =& v_1 x_{2 1} + v_2 x_{2 2} + 4 u x_{2 3},  \label{eq:h125eq2} \\
v^2=& v_1^2+v_2^2+4u^2, \label{eq:h125eq3}
 \end{align}
where we also added the constraint on the VEV's
to reproduce the weak scale (in our normalization, $v=246/\sqrt{2}\simeq174$GeV).
Note that since the linear combination of $v_1$
and $v_2$ is identical in \cref{eq:h125eq1,eq:h125eq2}, strict equality of the
two reduced couplings $\kappa_W^{h_{125}}$ and $\kappa_Z^{h_{125}}$ implies either $u=0$
or $x_{2 3} - \sqrt{2} x_{2 4}=0$. Barring the $u=0$
solution,  the two equations become degenerate
when
\begin{align}
 x_{2 3} - \sqrt{2} x_{2 4}=0.
 \label{eq:x23x24_relation}
\end{align}
Even when adding 
\cref{eq:h125eq3},
one has now some freedom in the values of the VEVs $v_1,v_2,u$ even for fixed $x_{2 i}$'s.

To proceed as already anticipated in \cref{sec:poss-sols}, we relax the above relation assuming the two reduced couplings not equal,
\begin{align}
\delta = \kappa_Z^{h_{125}} - \kappa_W^{h_{125}} \neq 0,
\end{align}
which leads to the consistency relation implied by \cref{eq:h125eq1,eq:h125eq2}, namely \cref{eq:ux_relation}. \Cref{eq:h125eq1,eq:h125eq2} degenerate then into one single equation:
\begin{align}
    \kappa_W^{h_{125}} = \delta + \frac{1}{v}\left(v_1 x_{2 1} + v_2 x_{2 2} + 4 \sqrt{2} u x_{2 4}\right). \label{eq:h125eq12}
\end{align}
Moreover, in order to treat the various Yukawa-type couplings within the same setup, we take $\kappa_d$ and $\kappa_u$ defined by \cref{eq:typeII_kappas} as input. 
Choosing
$|\kappa_d|\simeq1$ {\sl and} $|\kappa_u|\simeq1$ within the experimental error bars obviously means we are considering Type II (or Type Y and possibly Type X). On the other hand, 
taking $|\kappa_u|\simeq1$ within the experimental error bars for all quarks and charged leptons, and $|\kappa_d|$ outside the error bars, means implicitly that we consider Type I (and exclude Type X). 
Thus, taking  $\kappa_d$ defined by \cref{eq:typeII_kappas} as a free input allowed to be very different from one is interpreted as a parametrization of how much one deviates from Type II. 
Moreover, taking $\kappa_d$ and $\kappa_u$ as input has the merit of transforming \cref{eq:h125eq12} into a linear equation in $v_1^2, v_2^2$ which, coupled with \cref{eq:h125eq3}, give straightforwardly
$v_1^2$ and $v_2^2$ in terms of $\kappa_W^{h_{125}}, \delta, u, x_{2 4}, \kappa_d$ and $\kappa_u$. 

Since we stick to real-valued VEVs, the conditions $v_1^2>0$ and $v_2^2>0$ imply constraints that can be recast in terms of allowed intervals on $x_{24}$ that depend on the remaining input parameters $\kappa_W^{h_{125}}, \delta, \kappa_d, \kappa_u$, as well as $u$, the latter being  so far a free parameter. 
In addition, assuming all $x_{2j}$ real-valued, the normality condition $\sum_j x_{2j}^2  =1 $
should be required,  cf.\ \cref{eq:unitarity}. Using \cref{eq:ux_relation,eq:typeII_kappas} and the 
expressions of $v_1^2$ and $v_2^2$ in this condition leads to a two-fold solution for $x_{24}$ again in 
terms of $\kappa_W^{h_{125}}, \delta, \kappa_d, \kappa_u$ and $u$. The real-valuedness of $x_{24}$ is then found to imply a lower bound on 
$|u|$.

All the above constraints are controlled analytically. Finally, putting together any of the two-fold solutions for $x_{2 4}$ and its allowed domain that secures $v_1^2>0$ and $v_2^2>0$, leads to upper bounds on $|u|$. The value of $u$ can then be arbitrarily chosen within the determined upper and lower bounds on $|u|$ and only the solution fulfilling $|x_{2 4}|<1$ is consistently kept. The VEVs $v_1$ and $v_2$ are then uniquely fixed (up to global signs), as well as $x_{21},x_{22}$ and $x_{23}$ through \cref{eq:ux_relation,eq:typeII_kappas}.

In summary, providing the data for the couplings of the SM-like state $h_{125}$  to $W,Z$ and fermions and a consistent choice of the triplet VEV, fix entirely the doublets VEVs and the doublet/triplet content of $h_{125}$. The next step is to use plausible data for the other scalars, extracted from the experimental indications, in order to fully determine the neutral CP-even sector's couplings. We describe the procedure in the following section. 

\subsection{Information from $h_{95}$ and $H_{650}$ \label{app:A2}}
Since $v_1,v_2,u$ and the $x_{2i}$'s are now determined, one still needs to reconstruct the rest of the orthogonal matrix $\cal X$, with all its entries real-valued. The set of constraints \cref{eq:unitarity} 
boils down to reconstructing an orthonormal basis in 4 dimensions
starting from the known vector $(x_{2 i})$. To obtain, say, 
$(x_{1 i})$, one generates randomly three 4-dimensional vectors
(hopefully) forming with  $(x_{2 i})$ a linearly independent set, then construct, e.g. via the Gram-Schmidt method, an orthonormal basis for the 3-dimensional space orthogonal to $(x_{2 i})$. The vector $(x_{1 i})$ lives on the unit sphere centered at the origin in this space and thus depends on two angles. These angles are then fixed from the data of two reduced couplings, e.g. $\kappa_W^{h_{95}}$ and $\kappa_t^{h_{95}}$, upon use of 
\begin{align}
  \kappa_W^{h_{95}} =& \frac{1}{v}\left(
 v_1 x_{1 1} + v_2 x_{1 2} + 2 u (x_{1 3} + 
 \sqrt{2} x_{1 4})\right) \ {\rm and} \
 \kappa_t^{h_{95}} = \frac{v}{v_2} x_{12}.
 \label{eq:h95eq1}
\end{align} 
Knowing $(x_{1 i})$ and $(x_{2 i})$, the same Gram-Schmidt procedure is repeated to determine, say, the unit vector $(x_{4 i})$ which is parametrized by one angle in the 2-dimensional space orthogonal to $(x_{1 i})$ and $(x_{2 i})$. This angle is then determined from the data of, for example, $\kappa_W^{H_{650}}$
through the relation:
\begin{align}
  \kappa_W^{H_{650}} =& \frac{1}{v} \left(
 v_1 x_{4 1} + v_2 x_{4 2} + 2 u (x_{4 3} + 
 \sqrt{2} x_{4 4}) \right).
 \end{align}
 Finally, the remaining vector $(x_{3 i})$ is determined (up to a global sign), by applying Gram-Schmidt on $(x_{1 i}), (x_{2 i}), (x_{4 i})$ and a randomly chosen 4-dimensional vector.


\section{Getting the parameter space for $h_{95}$
\label{app:B}}
Let us briefly show how we got the allowed regions in the $\kappa_t$--$\kappa_W$ plane for $h_{95}$ from the 
decays $h_{95}\to \gamma\gamma$ and $h_{95}\to \tau^+\tau^-$. 

Suppose the interaction Lagrangian is 
\begin{equation}
{\cal L}_{\rm int} = -\frac{g m_t \kappa_t}{2m_W} \, \bar{t}t h + gm_W\kappa_W \, W_\mu^+ W^{\mu-}h 
- \frac{g m_H^2 \kappa_H}{m_W} \, H^+H^-h\,,
\end{equation}
where $h=h_{95}$, but it can be any other neutral scalar, even the SM Higgs. ${\cal L}_{\rm int}$ also includes
the contribution of some possible charged scalars in the theory. 

The decay width is given by \cite{Logan:2014jla}
\begin{equation}
\Gamma(h\to\gamma\gamma) = \frac{\alpha^2g^2}{1024\pi^3} \, \frac{m_h^3}{m_W^2}\, 
\left\vert \sum_i N_i \, e_i^2 \kappa_i \, F_i\right\vert^2\,,
\end{equation}
where $i$ can be 0, $\frac12$, or 1, depending on the spin of the particle in the loop, and $\kappa_0=\kappa_H$,
$\kappa_{1/2}=\kappa_t$, $\kappa_1=\kappa_W$. They modify the SM couplings. $N_i=1$ for $W$ and charged
scalars, and 3 for top; $e_i$ is the electric charge of the particle in the loop (1 for $W$ and $H^+$, $\frac23$ for top), 
and $\alpha=1/137$.

With the shorthand
\begin{equation} 
\tau= \frac{4m_i^2}{m_h^2}\,,\ \ \ f(\tau) = \left[ \sin^{-1}\left(1/\sqrt{\tau}\right)\right]^2\,,
\end{equation}
one has
\begin{eqnarray} 
F_1 &=& 2+3\tau+3\tau(2-\tau)\, f(\tau)\,,\nonumber\\
F_{1/2} &=& -2\tau\left[1+(1-\tau)\, f(\tau)\right]\,,\nonumber\\
F_0 &=& \tau\, \left[1-\tau\,f(\tau)\right]\,.
\end{eqnarray} 
Note that these expressions hold only if $2m_i > m_h$ so that there is no absorptive part. 

Suppose $h=h_{95}$, and we consider a Type-I 2HDM plus scalar triplets. $h_{95}$ can go only to $\tau^+\tau^-$ and
$b\bar{b}$ at tree-level; these are the only significant two-body channels that are open. So 
\begin{equation} 
{\rm Br} (h_{95} \to \tau^+\tau^-) \approx {\rm Br} (\phi \to \tau^+\tau^-)\,,
\end{equation} 
where $\phi$ is a 95 GeV scalar with exactly SM-like couplings. Now, $\sigma(pp\to h_{95}\to \tau^+\tau^-) 
= \sigma(pp \to h_{95})\times {\rm Br}(h_{95}\to \tau^+\tau^-)$. If $h_{95}$ is produced by ggF alone, $\mu_{\tau\tau}=
\kappa_t^2$; if it is produced by VBF alone, $\mu_{\tau\tau}=\kappa_W^2$. A better approximation, with a 
$90:10$ ratio for ggF and VBF modes of production, seems to be 
\begin{equation} 
\mu_{\tau\tau}=0.9\kappa_t^2 + 0.1\kappa_W^2\,.
\end{equation}
Without any charged Higgs, $\mu_{\gamma\gamma}$ is given by
\begin{equation} 
\mu_{\gamma\gamma} = (0.9\kappa_t^2+0.1\kappa_W^2)\, \frac{ {\rm Br}(h_{95}\to\gamma\gamma)}{ {\rm Br}
(\phi\to\gamma\gamma)} = (0.9\kappa_t^2+0.1\kappa_W^2)\, \frac{
\left\vert \kappa_W F_1 + \frac43 \kappa_t F_{1/2}\right\vert^2}{
\left\vert F_1 + \frac43 F_{1/2}\right\vert^2}
\end{equation}
Putting in all the values, this gives
\begin{equation} 
0.0295 (0.9\kappa_t^2+0.1\kappa_W^2) (-1.81\kappa_t+7.62\kappa_W)^2 = 0.33^{+0.19}_{-0.12}\,,
\end{equation}
while $\mu_{\tau\tau}$ gives $\sqrt{0.9 \kappa_t^2 + 0.1 \kappa_W^2} = 1.2\pm 0.5$. 

\section{Calculation of significances}
\label{app:significances}
The $p$-value for any signal is related the probability that the null hypothesis is acceptable; the smaller the $p$-value is,
the more strongly the null hypothesis ({\em i.e.}, no new physics) is rejected. This can also be converted to the 
significance quoted in terms of the standard deviation $\sigma$ for a normalized distribution. 

The experimentalists quote both the global and local significances, in terms of $\sigma$ or $p$-values. The local 
significance (LS) also includes the possibility of a random statistical fluctuation, called the look-elsewhere effect (LEE).
To eliminate LEE, we use the global significance (GS) numbers, with $\sigma_{\rm GS} < \sigma_{\rm LS}$. The ratio
of $\sigma_{\rm LS}:\sigma_{\rm GS}$ can be taken as the LEE factor. 

Suppose a new resonance is observed in two different channels, or maybe in the same channel by two different 
experiments. To start with, one may take the observations to be independent (this need not be correct, there may be 
correlations). A conservative, and according to us the correct, way is to combine the GS numbers, not the LS ones, at
least as long as the existence is confirmed. Once the existence is confirmed, there is no question of any LEE factor, 
and one may safely use the LS numbers for any further observations ({\em e.g.}, for any signal coming from the 
125 GeV Higgs resonance). The combined significance that one gets is naturally the most conservative one. A good
example is the significance of the $h_{95}$ signal as discussed in the text. 

An important question is how to combine two different $p$-values. Different methods exist with different statistical properties, see e.g. \cite{Heard:2018}. We will use two methods, the well-known Fisher's of obtaining the test statistic $\chi^2$\cite{Fisher:1934}, and the method of Ref.\ \cite{theiler} . We found that both 
methods give identical results.

While the $p$-value is related with the probability of rejecting the null hypothesis, it is not directly interpretable as a 
probability. Thus, the combination of two $p$-values, namely $p_1$ and $p_2$, is not simply $p_1p_2$. If this were
true, the product of a large number of statistically insignificant $p$-values could have resulted in a sufficiently small
$p$-value to discard the null hypothesis, which is obviously counter-intuitive. We describe here briefly the method of \cite{theiler}.

Suppose one has $n$ number of measurements with $p$-values denoted by $p_1$, $p_2 ..{..} p_n$ respectively, 
and let $p_* = p_1p_2 ..{..}p_n$, the product of all the $p$-values. Assuming the null hypothesis, let us 
compute the probability $R_n(\rho)$ such that $p_* < \rho$. Then $R_n(p_*)$ is the combined $p$-value. For $n=1$,
$R_1(\rho)=\rho$ by definition, and then one can use the recursion formula
\begin{equation}
R_n(\rho) = \rho + \int_{\rho}^1 R_{n-1}(\rho/p)\, dp\,,
\end{equation}
which leads to
\begin{equation}
p = p_1p_2 \left[1- \ln(p_1p_2)\right]
\end{equation}
for two measurements. The algorithm goes on {\em ad infinitum}.


\end{document}